\documentclass[12pt,a4paper]{article}
\pdfoutput=1
\usepackage{bm, amssymb,pifont,cancel, amsmath,comment,color}
\usepackage{here}
\usepackage{cite}
\usepackage{graphicx}
\usepackage{subfigure}

\makeatletter

\newcommand{\be}{\begin{equation}}
\newcommand{\ee}{\end{equation}}
\newcommand{\bea}{\begin{eqnarray}}
\newcommand{\eea}{\end{eqnarray}}

\def \Mpl {M_{\text{Pl}}}

\begin{document}

\begin{titlepage}

\begin{centering}
\vspace{1cm}
{\Large {\bf Reheating and Dark Matter Freeze-in \vspace{.15cm} \\ in the Higgs-$R^2$ Inflation Model}} \\

\vspace{1.5cm}

{\bf  Shuntaro Aoki$^{\dagger}$, Hyun Min Lee$^{*}$, Adriana G. Menkara$^\star$,  \vspace{.15cm} \\
and Kimiko Yamashita$^{\sharp}$}
%\\

\vspace{.5cm}

{\it  Department of Physics, Chung-Ang University, Seoul 06974, Korea.}

\end{centering}

\vspace{2cm}

\begin{abstract}
\noindent
We study the post-inflationary dynamics for reheating and freeze-in dark matter in the Higgs-$R^2$ inflation model. Taking the perturbative approach for reheating, we determine the evolution of the temperature for radiation bath produced during reheating and determine the maximum and reheating temperatures of the Universe. Adopting a singlet scalar dark matter with a conformal non-minimal coupling and a vanishing Higgs-portal coupling, we discuss the freeze-in production of dark matter both from the non-thermal scattering during reheating and the thermal scattering after reheating. We find that  thermal scattering is dominant for dark matter production in our model due to the high reheating temperature. The reheating temperature in our model is determined dominantly by the Higgs condensate to be up to about $10^{14}\,{\rm GeV}$ and dark matter with masses up to about $10^9\,{\rm GeV}$ can be produced with a correct relic density.

\end{abstract}

\vspace{4cm}

\begin{flushleft} 
$^\dagger$Email:  shuntaro@cau.ac.kr  \\
$^*$Email: hminlee@cau.ac.kr \\
$^\star$Email: amenkara@cau.ac.kr \\
$^\sharp$Email: kimikoy@cau.ac.kr
\end{flushleft}

\end{titlepage}

\tableofcontents
\vspace{35pt}
\hrule

%%%%%%%%%%%%%%%%%%%%%%%%%%%%%%%%%%%%%%%%%%%%%%%%%%%%%%%%%%%%%%%%%%%%%%%%%%%%%%%%%%%%%%%%%%%%%%%%%%%%%%%%%%%%%%%%%%%%%%%%%%%%%%%%%%%%%%%%%%%%%%%%%%%%%%%%%%%%%%%%%%%%%%%%%%%%%%%%%%%%%%%%%%%%%%%%%%%%%%%%%%%%%%%%%%%%%%%%%%%%%%%%%%%%%%%%%%%%%%%%%%%%%%%%%%%%%%%%%%%%%%%%%%%%%%%%%%%%%%%%%%%%%%%%%%%%%%%%%%%%%%%%%%%%%%%%%%%%%%%%%%%%%%%%%%%%%%%%%%%%%%%%%
\section{Introduction}\label{intro}

%inflation
Higgs inflation~\cite{Bezrukov:2007ep} has been drawing a lot of attention for recent years due to the fact that the Higgs boson in the Standard Model (SM), which was discovered at the Large Hadron Collider, can play a role for slow-roll inflation in the early Universe. Thus, it provides a testing ground for inflationary scenarios by the interplay between the Higgs data at small scales and the inflationary observables at large scales. The original proposal for Higgs inflation, however, has a unitarity problem, because a large non-minimal coupling is required to match the anisotropies of Cosmic Microwave Background (CMB) and it leads to a premature violation of unitarity of order the Hubble scale during inflation \cite{Burgess:2009ea,Burgess:2010zq,Barbon:2009ya,Hertzberg:2010dc}. There are proposals to resolving the unitarity problem beyond the Higgs inflation by adding a new degree of freedom coupled to the Higgs boson \cite{Giudice:2010ka,Barbon:2015fla}. Among the proposed solutions is the ultra-violet (UV) completion of linear sigma model type \cite{Giudice:2010ka}, extending the global symmetry of the Lagrangian in the Higgs-sigma field space. The extension of the Higgs inflation with an $R^2$ term has been identified as a linear sigma model \cite{Ema:2020zvg,Lee:2021dgi,Aoki:2021aph}, so it is amusing to make a dual field-theory interpretation of the gravitational couplings in this context.

%reheating
The Universe would have been empty after inflation unless there is a mechanism for transferring the inflation energy to a hot thermal plasma. Thus, the period of reheating is necessary to make a smooth transition from inflation to hot Big Bang Universe \cite{Traschen:1990sw,Kofman:1994rk,Kofman:1997yn}. However, reheating depends on the couplings between the inflaton and the SM particles, which are completely unknown in inflation models with a singlet inflaton. It is remarkable that if there is a delay in the completion of reheating due to small inflaton couplings, the detailed evolution of reheating dynamics, such as the equation of state and the reheating temperature, etc, could alter the inflationary predictions \cite{Choi:2016eif}. On the other hand, in Higgs inflation and its UV complete models, the inflaton couplings to the rest of the SM particles are fixed by the gauge symmetry of the SM and the new symmetry restoring the unitarity.  Thus, it is important to study the reheating  dynamics concretely in these models and check the consistency for inflation.

%inflation + dark matter
Not only hot thermal plasma with visible particles but also dark matter and dark energy are necessary ingredients for the success of standard cosmology. There is a variety of evidence for dark matter from galaxy rotation curves, gravitational lensing, CMB, Bullet cluster, etc, but we don't know the origin of dark matter in particle physics. Depending on the interactions between dark matter and the SM particles, we can determine the dark matter abundance at present and make a strategy for designing direct and indirect detection experiments for dark matter. Given that there is no convincing direct evidence for dark matter, it may be the case that dark matter is sequestered from the SM and it may interact with the SM very feebly~\cite{Hall:2009bx,Chu:2011be,Bernal:2017kxu,Choi:2018lxt}.

%paper
In this article, we investigate the reheating dynamics and the dark matter freeze-in process in the extension of Higgs inflation with an $R^2$ term. For inflation and reheating discussion, we take the linear-sigma model frame where the conformal symmetry for gravitational couplings is manifest and scalar fields have canonical kinetic terms. We introduce a singlet scalar dark matter in our model with a near-conformal non-minimal coupling to gravity and a vanishing small Higgs-portal coupling. There was a study on the production of primordial black holes as dark matter in the Higgs-$R^2$ inflation~\cite{Cheong:2019vzl,Cheong:2021vdb}.

Based on the perturbative analysis for reheating, we determine the evolution of the inflaton condensates and the temperature of the Universe during reheating. Using the results for reheating, we consider the freeze-in production of scalar dark matter by the non-thermal inflaton scattering and the thermal scattering between SM particles and find the parameter space for explaining the correct relic density for dark matter. We also take into account the gravitational production of dark matter via massless graviton.

%organize
The paper is organized as follows. 
We begin with the setup for the Higgs inflation model amended with an $R^2$ term and recast it into a linear-sigma model dual Lagrangian. We discuss the  main features of inflationary predictions and the perturbativity conditions during inflation. Next we focus on the perturbative reheating by using the Boltzmann equations for inflaton and radiation energy densities and determine the evolution of the temperature during reheating. We continue to introduce a singlet scalar dark matter in our model with a general non-minimal coupling and a Higgs-portal coupling and discuss the dark matter production during and after reheating. Finally, conclusions are drawn. There is an appendix dealing with the details on thermal scattering rates for dark matter production in our model. 

Throughout the paper, we use the mostly plus convention $(-,+,+,+)$ for the metric.

%%%%%%%%%%%%%%%%%%%%%%%%%%%%%%%%%%%%%%%%%%%%%%%%%%%%%%%%%%
\section{Higgs-$R^2$ inflation}

We first introduce the setup for the Higgs-$R^2$ inflation model and discuss the effective inflaton potential and its predictions for inflationary observables. We also show the constraints on the model parameters from perturbativity and CMB measurements.

%%%%%%%%%%%%%%%%%%%%%%%%%%%%%%%%%%%%%%%%%%%%%%%%%%%%%%%%%%

\subsection{The model}
Introducing the non-minimal coupling for Higgs fields in the SM and the $R^2$ term beyond the Einstein gravity, we begin with the corresponding Lagrangian~\cite{Salvio:2015kka,Salvio:2016vxi,Ema:2017rqn,Gorbunov:2018llf,Gundhi:2018wyz}, as follows,
\bea
\mathcal{L} / \sqrt{-g_J}=\frac{1}{2}(\Mpl^2+\xi \hat{h}^2)R_J -\frac{1}{2}(\partial_{\mu} \hat{h})^{2}-\frac{\lambda}{4}\hat{h}^4+\alpha R_J^2 \label{L_J}
\eea
where $g_{J \mu\nu}$ and $R_{J}$ are the spacetime metric and the Ricci scalar in Jordan frame, respectively, $\Mpl=2.4\times 10^{18}$ GeV is the reduced Planck mass, and $\hat{h}$ is  the Standard Model Higgs boson in unitary gauge. We omit the Higgs mass parameter during inflation and reheating. We note that $\xi$ and $\lambda$ are the non-minimal coupling and the quartic coupling for the Higgs boson, respectively, and $\alpha$ is the coefficient of the $R^2$ term. The Higgs-$R^2$ model with Eq.~$\eqref{L_J}$ provides a unitary completion of the original Higgs inflation up to the Planck scale and it also explains the CMB data well.

Following the discussion in Refs.~\cite{Ema:2020zvg,Ema:2020evi,Lee:2021dgi,Aoki:2021aph}, we change the original frame in Eq.~$\eqref{L_J}$ to a new frame where the unitarity up to the Planck scale is manifest. 
To do so, we first introduce an auxiliary field $\hat{\chi}$ instead of the $R^2$ term, in the following,
\begin{align}
\mathcal{L} / \sqrt{-g_J}=\frac{1}{2}(\Mpl^2+\xi \hat{h}^2+4\alpha\hat{\chi})R_J -\frac{1}{2}(\partial_{\mu} \hat{h})^{2}-\frac{\lambda}{4}\hat{h}^4-\alpha\hat{\chi}^2.
\end{align}
Then, we can check that the original Lagrangian with the $R^2$ term in Eq.~$\eqref{L_J}$ is reproduced after $\hat{\chi}$ is integrated out. 
Next we perform a conformal transformation with following field redefinition\footnote{The new frame is called the {\it{Linear-sigma}} frame. As shown below, the introduction of $\sigma$-field linearize the original Higgs inflation model in the new frame, which is analogous to the $\sigma$-field in the linear sigma model.} ,
\begin{align}
g_{J\mu\nu}=\Delta^{ -2}g_{L\mu\nu}, \ \ \hat{h}= \Delta h, \ \ \hat{\chi}= \Delta^2 \chi,  \label{g_Jtog_L} 
\end{align}
with
\begin{align}
\Delta^{ -2}=\left(1+\frac{\sigma}{\sqrt{6}\Mpl}\right)^{2}, \label{def_Delta}  
\end{align}
and the $\sigma$ field being subject to the following constraint,
\bea
 \left(1+\frac{\sigma}{\sqrt{6}\Mpl}\right)^{2}+\xi \frac{h^2}{\Mpl^2}+4\alpha \frac{\chi}{\Mpl^2}=1-\frac{h^2}{6\Mpl^2}-\frac{\sigma^2}{6\Mpl^2}.
\eea
Thus, we have changed the fundamental variable from $\chi$ to $\sigma$ in Eq.~$\eqref{def_Delta}$. As a result, the Lagrangian is given in terms of $(h,\sigma)$  by
\bea
\mathcal{L} / \sqrt{-g_L}&=&\frac{\Mpl^2}{2}\left(1-\frac{h^{2}}{6\Mpl^2} -\frac{\sigma^{2}}{6\Mpl^2} \right) R_L-\frac{1}{2}\left(\partial_{\mu} \sigma\right)^{2} -\frac{1}{2}\left(\partial_{\mu} h\right)^{2}-\frac{\lambda}{4}h^4 \nonumber \\
&&-\frac{\kappa}{4}\left[\sigma(\sigma+\sqrt{6}\Mpl)+3\left(\xi+\frac{1}{6}\right)h^2\right]^2,\label{L_L} \label{linearsigma}
\eea
with $\kappa\equiv 1/(36\alpha)$. In the new frame, $\sigma$ and $h$ conformally couples to the Ricci scalar and their kinetic terms are canonically normalized (i.e., the field target space is flat), so  unitarity and perturbativity are manifest. It is remarkable that the running Higgs quartic coupling is corrected by the Higgs non-minimal coupling above the sigma scalar threshold to $\lambda_{\rm{eff}}=\lambda+9\kappa\big(\xi+\frac{1}{6})^2$, so the stability of the electroweak vacuum can be guaranteed due to the tree-level shift in the Higgs quartic coupling \cite{Elias-Miro:2012eoi}.

\subsection{Effective inflaton potential}

We discuss the  inflationary prediction of the Higgs-$R^2$ inflation model in the Einstein frame. 

Making a Weyl transformation with
\begin{align}
g_{L\mu\nu}=\Omega^{ -2}g_{E\mu\nu} ,   \qquad
\Omega^{ 2}=1-\frac{h^{2}}{6\Mpl^2}-\frac{\sigma^{2}}{6\Mpl^2} ,    
\end{align}
we recast Eq.~$\eqref{L_L}$ into the Einstein frame Lagrangian,
\bea
\mathcal{L} / \sqrt{-g_E}&=&\frac{\Mpl^2}{2}R_E -\frac{1}{2\Omega^4}\left(1-\frac{h^{2}}{6\Mpl^2} \right)\left(\partial_{\mu} \sigma\right)^{2}-\frac{1}{2\Omega^4}\left(1-\frac{\sigma^{2}}{6\Mpl^2} \right)\left(\partial_{\mu} h\right)^{2} \nonumber \\
&&-\frac{h \sigma}{6\Mpl^2\Omega^4}  \partial_{\mu} h \partial^{\mu} \sigma-V, \label{L_E}  
\eea
where 
\begin{align}
V=  \frac{1}{\Omega^4}\left[\frac{1}{4} \kappa\left(\sigma(\sigma+\sqrt{6}\Mpl)+3\left(\xi+\frac{1}{6}\right) h^{2}\right)^{2}+\frac{1}{4} \lambda h^{4}\right].  \label{V}
\end{align}
In the following discussion, we omit $``E"$ for the Einstein metric. 

During inflation, $h$ has a large mass much greater than the Hubble scale $H$~\cite{Ema:2017rqn} (see also Eq.~$\eqref{h_mass_inf}$), so that it can be integrated out. It turns out that $\frac{dV}{dh}=0$ leads to a nonzero VEV of $h$ \cite{Lee:2021dgi}, as follows,
\begin{align}
h^{2}=\frac{\kappa \sigma(\sigma+\sqrt{6}\Mpl)\left(\sigma-3\left(\xi+\frac{1}{6}\right)(\sigma-\sqrt{6}\Mpl)\right)}{\lambda(\sigma-\sqrt{6}\Mpl)-3 \kappa\left(\xi+\frac{1}{6}\right)\left(\sigma-3\left(\xi+\frac{1}{6}\right)(\sigma-\sqrt{6}\Mpl)\right)}.    \label{happ}
\end{align}
Then, inserting Eq.~(\ref{happ}) back to the Lagrangian~$\eqref{L_E}$, we obtain the effective Lagrangian for~$\sigma$,
\begin{align}
\mathcal{L}_{\mathrm{eff}}/\sqrt{-g}=\frac{\Mpl^2}{2} R-\frac{\left(\partial_{\mu} \sigma\right)^{2}}{2\left(1-\frac{\sigma^{2}} { 6\Mpl^2}\right)^{2}}-V_{\mathrm{eff}}(\sigma),   
\end{align}
where the effective inflaton potential is given by
\begin{align}
V_{\mathrm{eff}}(\sigma)=9 \lambda \kappa \Mpl^4 \sigma^{2}\left[\lambda(\sigma-\sqrt{6}\Mpl)^{2}+\kappa\left(\sigma-3\left(\xi+\frac{1}{6}\right)(\sigma-\sqrt{6}\Mpl)\right)^{2}\right]^{-1}.    
\end{align}
In terms of the canonical field $\phi$ defined through
\begin{align}
\sigma/\Mpl=-\sqrt{6} \tanh \left(\frac{\phi}{\sqrt{6}\Mpl}\right),    
\end{align}
we express the effective inflaton potential  \cite{Lee:2021dgi} as
\begin{align}
\nonumber V_{\mathrm{eff}}(\phi)=&\ \frac{9 \kappa \Mpl^4}{4}\left(1-e^{-\frac{2 \phi}{\sqrt{6} M_{\mathrm{Pl}}}}\right)^{2}\left[1+\frac{\kappa}{4 \lambda}\left(6 \xi+1+e^{-\frac{2 \phi}{\sqrt{6} M_{\mathrm{Pl}}}}\right)^{2}\right]^{-1}\\
\simeq &\ V_I\left(1-\frac{2\lambda+\kappa (3\xi+1)(6\xi+1)}{\lambda+9\kappa \big(\xi+\frac{1}{6}\big)^2}\, \cdot e^{-\frac{2 \phi}{\sqrt{6} M_{\mathrm{Pl}}}}+\cdots \right),\label{EV}    
\end{align}
with 
\bea
V_I\equiv \frac{9\kappa \lambda \Mpl^4}{4\left(\lambda+9\kappa \big(\xi+\frac{1}{6}\big)^2\right)}. 
\eea
We note that the scalar potential is very flat for $\phi/\Mpl\gg 1$, and it unifies the $R^2$ inflation and 
the Higgs inflation: $R^2$-like (or Higgs-like) inflation can be realized for $9 \kappa \xi^{2} \ll \lambda$ (or $9 \kappa \xi^{2} \gg \lambda$).

We remark that the decoupling condition for the Higgs in Eq.~(\ref{happ}) takes an approximate form during inflation,
\bea
h^2\simeq \frac{72\kappa\xi\Mpl^2}{2\lambda+3\kappa\xi (6\xi+1)}\, e^{-\frac{2 \phi}{\sqrt{6} M_{\mathrm{Pl}}}}\equiv A\Mpl^2\,e^{-\frac{2 \phi}{\sqrt{6} M_{\mathrm{Pl}}}}.
\eea
So, for $A>0$, the Higgs field is stabilized at a nonzero sigma-dependent background value during inflation, and a positive squared mass for the canonically normalized Higgs boson is obtained during inflation as
\bea
m^2_h=12\xi\bigg(2+\frac{3\kappa\xi}{\lambda}\, (1+6\xi) \bigg)H^2_I =\frac{864\kappa \xi^2}{A\lambda}\, H^2_I, \label{h_mass_inf}
\eea
where $H_I\simeq \sqrt{\frac{V_I}{3M^2_{\rm Pl}}}$ is the Hubble scale during inflation. For $R^2$-like (or Higgs-like) inflation, we obtain $m^2_h\simeq 24\xi H^2_I$ (or $m^2_h\simeq 216\kappa \xi^3/\lambda\, H^2_I$).
Thus, in order to safely decouple the Higgs field during inflation, we need to take $m_h\gg H_I$, requiring $\xi\gg 1$. 

On the other hand, for $A<0$,  the Higgs field could not be stabilized at a nonzero value, so instead we need to take $h=0$ during inflation in this case. As a result, the inflaton potential becomes the one for Starobinsky model, as follows,
\bea
V_{\rm eff}(\phi)\simeq \frac{9\kappa M^4_{\rm Pl}}{4}\, \left(1-e^{- \frac{2\phi}{\sqrt{6}\Mpl}}\right)^{2},
\eea
with the effective Higgs mass given by
\bea
m^2_h=-12 \xi H^2_I.
\eea
Thus, for $\xi<0$, the Higgs direction is stable during inflation. But, for $\xi<0$, the graviton kinetic term in the original Lagrangian with the $R^2$ term in Eq.~(\ref{L_J}) could have a wrong sign beyond  a certain Higgs field value, so we don't consider the possibility with $\xi<0$ in this work.

\subsection{Inflationary observables and perturbativity}

The CMB normalization of the scalar power spectrum gives a relation in the parameters (see Ref.~\cite{Lee:2021dgi} for details),
\begin{align}
\frac{ \lambda+9\kappa \big(\xi+\frac{1}{6}\big)^2}{\kappa \lambda}=2.25\times 10^{10}. \label{CMB}
\end{align}
The slow-roll parameters are given by
\bea
\epsilon&=&\frac{\Mpl^2}{2}\left(\frac{1}{V_{\rm eff}} \frac{d V_{\rm eff}}{d \phi}\right)^{2} = \frac{1}{3} \frac{\left(2 \lambda+ \kappa (1+3\xi)(1+6 \xi)\right)^{2}}{\left(\lambda+9 \kappa \big(\xi+\frac{1}{6}\big)^{2}\right)^{2}} e^{-\frac{4 \phi}{  \sqrt{6}\Mpl}}, \\
\eta &=&\frac{\Mpl^2}{V_{\rm eff}}\frac{d^2V_{\rm eff}}{d\phi^2}  =-\frac{2}{3} \, \cdot \frac{2\lambda+\kappa(1+3\xi)(1+6\xi)}{\lambda+9\kappa\big(\xi+\frac{1}{6}\big)^2}\, e^{-\frac{2 \phi}{  \sqrt{6}\Mpl}}  \nonumber \\
&&\qquad\qquad\quad\,\,+ \frac{2\kappa}{3}\, \cdot\frac{(\lambda+12\lambda\xi+\kappa(1+3\xi)(1+6\xi)^2)}{\left(\lambda+9\kappa\big(\xi+\frac{1}{6}\big)^2\right)^2}\, e^{-\frac{4 \phi}{  \sqrt{6}\Mpl}}.
\eea
Then, the spectral index and the tensor-to-scalar ratio  are given in terms of the number of e-folding $N$ by
\bea
n_s&=&1-6\epsilon_* +2\eta_* \nonumber \\
&=&1-\frac{2}{N}-\frac{9}{2 N^{2}}+\frac{3 \kappa}{N^{2}} \frac{\left(\lambda+12 \lambda \xi+\kappa (1+3\xi)(1+6 \xi)^2\right)}{\left(2 \lambda+\kappa(1+3 \xi)(1+6 \xi)\right)^{2}}, \label{spectral} \\
r&=&16\epsilon_*=\frac{12}{N^2}, \label{tensor}
\eea
where $\epsilon_*,\eta_*$ are the slow-roll parameters evaluated at the horizon exit. The inflationary predictions for $N=50-60$ are consistent with the Planck result~\cite{Planck:2018jri}.

In the case with non-instantaneous reheating, we get the number of efoldings required to solve the horizon problem, as follows \cite{Choi:2016eif},
\bea
N=61.1 +\Delta N-\ln \bigg(\frac{V^{1/4}_{\rm end}}{H_k} \bigg)-\frac{1}{12} \ln \bigg(\frac{g_{\rm reh}}{106.75}\bigg), \label{Nefolds}
\eea
where the contribution from the delayed reheating is given by
\bea
\Delta N=\frac{1}{12} \bigg(\frac{3w-1}{w+1}\bigg) \,\ln\bigg(\frac{45 V_{\rm end}}{\pi^2 g_{\rm reh} T^4_{\rm reh}}\bigg).
\eea
Here, $V_{\rm end}$ is the inflation energy at the end of inflation, $H_k$ is the Hubble parameter evaluated at the horizon exit for the Planck pivot scale, $k=0.05\,{\rm Mpc}^{-1}$, and $g_{\rm reh}, T_{\rm reh}$ are the number of massless degrees of freedom and the reheating temperature at reheating completion, respectively, and $w$ is the averaged equation of state during reheating.  

Inflation ends when $\epsilon=1$. Then, we read off the inflaton field value at the end of inflation, $\phi_e$, as
\begin{align}
\phi_e/\Mpl=\frac{\sqrt{6}}{4}\log \left(\frac{\left(2 \lambda+\kappa (1+3\xi)(1+6 \xi)\right)^{2}}{3\left(\lambda+9 \kappa \big(\xi+\frac{1}{6}\big)^{2}\right)^{2}} \right).    
\end{align}
In either $R^2$-like or Higgs-like inflations, the argument of the logarithm is roughly estimated as $4/3$, which leads to
\begin{align}
\phi_e/\Mpl\simeq 0.18, \label{chi_e}
\end{align}
or $\sigma_e/\Mpl\simeq -0.18$.
Then, the inflaton field value at the end of inflation sets the initial condition for inflaton condensates at the onset of oscillations.

\begin{figure}[t]

  \begin{center}
   \includegraphics[width=90mm]{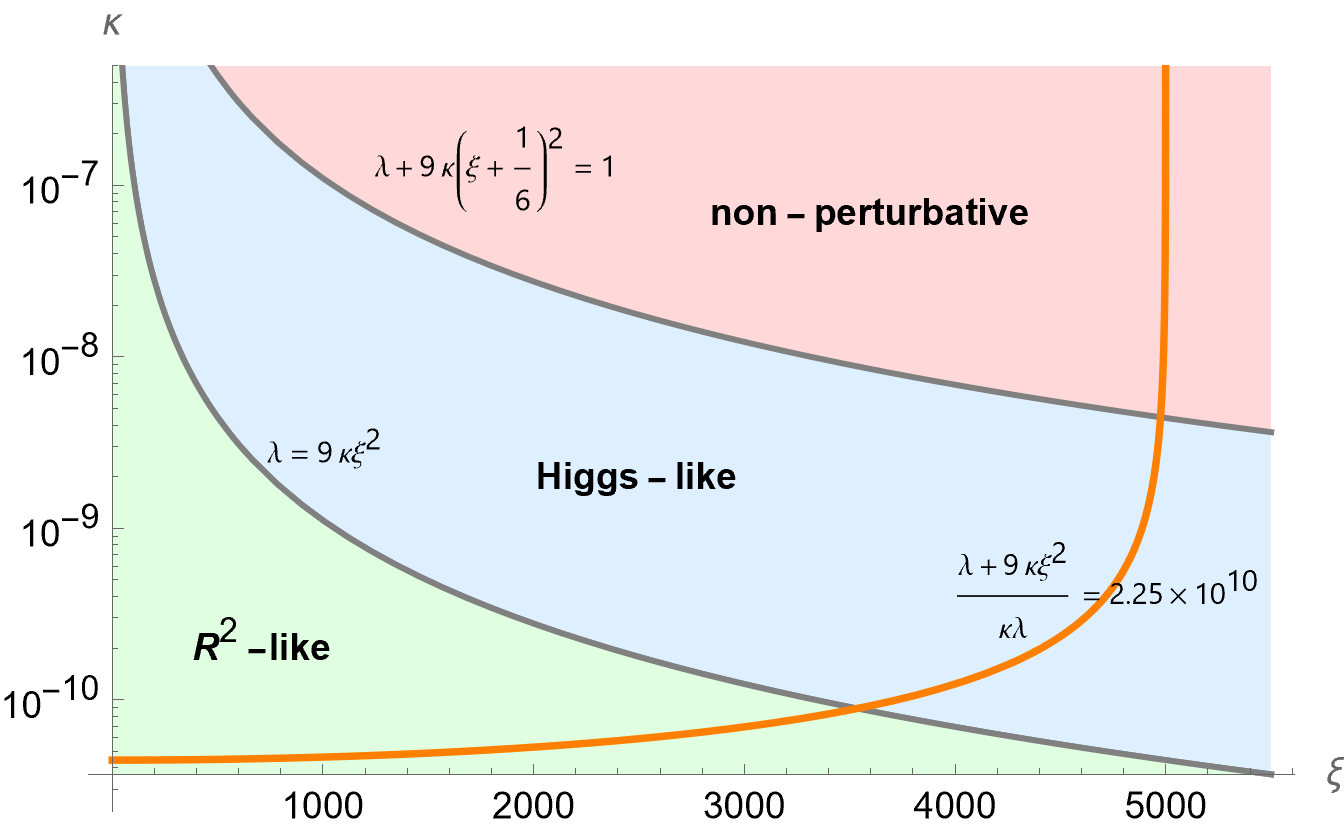}
  \end{center}
\caption{Consistent inflation for the parameter space in $(\xi,\kappa)$. The red region shows the strong coupling regime. The blue and green regions correspond to the Higgs-like inflation (with $9 \kappa \xi^{2} > \lambda$) and the $R^2$-like inflation (with $9 \kappa \xi^{2} < \lambda$), respectively. The orange line satisfies the CMB normalization~$\eqref{CMB}$. We set $\lambda=0.01$. }
  \label{fig:kappaxi}
 
\end{figure}

From the Lagrangian (\ref{linearsigma}) in the linear sigma-model frame, it is obvious that there is no unitarity violation up to the Planck scale, provided that the following perturbativity conditions are satisfied:
 \begin{align}
\kappa \lesssim 1, \quad \lambda_{\rm eff}\equiv\lambda+9 \kappa\left(\xi+\frac{1}{6}\right)^{2} \lesssim 1, \quad 6 \kappa\left(\xi+\frac{1}{6}\right) \lesssim 1.   \label{perturbativity} 
\end{align}
It is remarkable that the large non-minimal coupling $\xi$ in the original Higgs inflation accompanies with a new parameter~$\kappa$ (the inverse coefficient of $R^2$-term) and their product could be small or order one. 
Thus, now we consider the constraints on parameters from the perturbativity conditions~$\eqref{perturbativity}$ and the CMB constraint~$\eqref{CMB}$. In Fig.~\ref{fig:kappaxi}, we show the allowed parameter space of $(\xi,\kappa)$ with $\lambda=0.01$ fixed\footnote{The value of the Higgs quartic coupling $\lambda$ at inflation scales depends on the SM parameters, such as top quark mass and Higgs mass, through the renormalization group equations. But, for concreteness, we will choose $\lambda=0.01$ at inflation in the current section and in the following discussion on reheating.}. The red region does not satisfy the perturbativity conditions, which means the strong coupling regime. The perturbative regions are further divided by Higgs-like (blue) and $R^2$-like (green) situations. The orange line corresponds to the CMB constraint~$\eqref{CMB}$. In the following sections, we will discuss some phenomenological aspects of the Higgs-$R^2$ model while
keeping these conditions in mind.

%%%%%%%%%%%%%%%%%%%%%%%%%%%%%%%%%%%%%%%%%%%%%%%%%%%%%%%%%%

\section{Perturbative reheating}

We consider the perturbative reheating in the Higgs-$R^2$ model. To this end, we solve the Boltzmann equations for the energy densities for inflaton and radiation in the presence of inflaton decay rates, and obtain the time evolution of the inflaton condensates and the radiation energy density during reheating. 
Using the numerical results, we identify the maximum and reheating temperatures of the Universe during reheating. 
Our approach for reheating can be compared to the oscillation condensate with dissipation in non-equilibrium thermodynamics \cite{Ai:2021gtg,Wang:2022mvv}.

\subsection{Boltzmann equations during reheating}

In order to discuss the reheating process, we consider the system of dynamical equations, composed of the Boltzmann equations for the inflaton condensates and the radiation energy as well as the Friedmann equation, as follows,
\bea
&&\ddot{\sigma}+\frac{\sigma}{3\Omega^{2}\Mpl^2}\dot{\sigma}^{2}+\frac{h}{3\Omega^{2}\Mpl^2} \dot{\sigma} \dot{h}+(3H+\Gamma_{\sigma_0})\dot{\sigma} \nonumber \\
&&+\frac{2 \sigma}{3\Omega^{2}\Mpl^2} U+\frac{1}{\Omega^{2}}\left(1-\frac{\sigma^{2}}{6\Mpl^2}\right) U_{\sigma}-\frac{h \sigma}{6\Omega^{2}\Mpl^2} U_{h}=0,\label{EOM_sigma}  \\
&&\ddot{h}+\frac{h}{3\Omega^{2}\Mpl^2}\dot{h}^{2}+\frac{\sigma}{3\Omega^{2}\Mpl^2} \dot{\sigma} \dot{h}+(3H+\Gamma_{h_{\rm{osc}}})\dot{h}
\nonumber \\
&&+\frac{2 h}{3\Omega^{2}\Mpl^2} U+\frac{1}{\Omega^{2}}\left(1-\frac{h^{2}}{6\Mpl^2}\right) U_{h}-\frac{h \sigma}{6\Omega^{2}\Mpl^2} U_{\sigma}=0, \\
&&\dot{\rho}_r+4H\rho_r-\frac{\Gamma_{\sigma_0}}{\Omega^4}\left[\left(1-\frac{h^{2}}{6\Mpl^2}\right)\dot{\sigma}^2+\frac{h\sigma}{6\Mpl^2}\dot{\sigma}\dot{h}\right] \nonumber \\
&&-\frac{\Gamma_{h_{\rm{osc}}}}{\Omega^4}\left[\left(1-\frac{\sigma^{2}}{6\Mpl^2}\right)\dot{h}^2+\frac{h\sigma}{6\Mpl^2}\dot{\sigma}\dot{h}\right]=0,\label{rho_r} \\
&&3 H^{2}\Mpl^2 =\rho_{\sigma+h}+\rho_r \label{Friedmann}
\eea
where $U\equiv V\Omega^4$ and the subscripts in $U_\sigma$, etc, denote the derivatives with respect to corresponding fields.
Here, $\rho_{\sigma+h}$ is the total energy density for $\sigma$ and $h$, given by
\begin{align}
\rho_{\sigma+h}\equiv \frac{1}{2\Omega^4}\left[\left(1-\frac{h^{2}}{6\Mpl^2}\right) \dot{\sigma}^{2}+\left(1-\frac{\sigma^{2}}{6\Mpl^2}\right) \dot{h}^{2}+\frac{h \sigma}{3\Mpl^2} \dot{\sigma} \dot{h}+2U\right], 
\end{align}
$\rho_r$ is a radiation energy density, and $\Gamma_{\sigma_0}$ and $\Gamma_{h_{\rm{osc}}}$ are the decay rates of the sigma and Higgs condensates, which will be derived shortly. 

We consider the equation of state parameter $w$ as an important quantity to see the evolution of the Universe after inflation,
\begin{align}
w=\frac{p_{\sigma+h}+p_{r}}{\rho_{\sigma+h}+\rho_{r}}=\frac{p_{\sigma+h}+\rho_{r}/3}{\rho_{\sigma+h}+\rho_{r}},    
\end{align}
where $p_{\sigma+h}$ is the pressure for the inflaton condensates,
\begin{align}
p_{\sigma+h}\equiv \frac{1}{2\Omega^4}\left[\left(1-\frac{h^{2}}{6\Mpl^2}\right) \dot{\sigma}^{2}+\left(1-\frac{\sigma^{2}}{6\Mpl^2}\right) \dot{h}^{2}+\frac{h \sigma}{3\Mpl^2} \dot{\sigma} \dot{h}-2U\right],
\end{align}
and $p_r=\rho_r/3$ is the pressure for the radiation.

\subsection{Background field evolution after inflation}\label{BG_after_inflation}

After inflation, the sigma field starts to oscillate around the potential minimum. In the mean time, the Higgs field is released from the initial background field value set by Eq.~$\eqref{happ}$, and it also starts to oscillate between the broken and symmetric phases of electroweak symmetry, depending on the sign of $\sigma$, and it contains a rapidly oscillating part~\cite{Bezrukov:2019ylq,He:2020ivk,Bezrukov:2020txg,He:2020qcb,He:2018mgb}. 

We denote the background evolution of $\sigma$ by $\sigma_0$ and divide the background evolution of $h $ into a slowly oscillating part $h_0$ related to $\sigma_0$ and a rapidly oscillating part $h_{\rm{osc}}$ \cite{Fan:2019udt,He:2020qcb}, as follows,
\bea
\sigma(t)&=&\sigma_0, \\
h(t)&=&  h_0(\sigma_0)+h_{\rm{osc}}(t),  \label{h_0} 
\eea
where the inflation condition in Eq.~(\ref{happ}) relates the Higgs condensate to the sigma condensate by
\bea
(h_0(\sigma_0))^2&=& \frac{\kappa \sigma_0(\sigma_0+\sqrt{6}\Mpl)\left(\sigma_0-3\tilde{\xi}(\sigma_0-\sqrt{6}\Mpl)\right)}{\lambda(\sigma_0-\sqrt{6}\Mpl)-3 \kappa\tilde{\xi}\left(\sigma_0-3\tilde{\xi}(\sigma_0-\sqrt{6}\Mpl)\right)} \nonumber \\
&\simeq& -\frac{3\sqrt{6}\kappa\tilde{\xi}}{\lambda+9\kappa\tilde{\xi}^2}\Mpl\sigma_0,   \label{hv}   
\eea
for $\sigma_0<0$, and $h_0=0$ for $\sigma_0>0$. For $\sigma_0<0$, we used  $|\sigma_0|/\Mpl\ll 1$. Henceforth, we take a simpler notation for the Higgs non-minimal coupling to
\begin{align}
\tilde{\xi}\equiv \xi+1/6.    
\end{align}
It turns out that the above relation in Eq.~(\ref{hv}) is a good approximation during reheating for $\tilde{\xi}\gtrsim 100$. 
The behavior can be understood from the cubic coupling of type, $\kappa \tilde{\xi}\sigma h^2$, in the scalar potential $\eqref{V}$, because the negative mass term for $h$ with $\sigma<0$ develops the non-zero VEV. On the other hand, when $\sigma>0$, we find that the $h_0$ part of the Higgs condensate goes to zero, which is now a stable minimum. We confirm this behavior by solving the equations of motion numerically as below.

\begin{figure}[t]
 \begin{minipage}{0.5\hsize}
  \begin{center}
   \includegraphics[width=70mm]{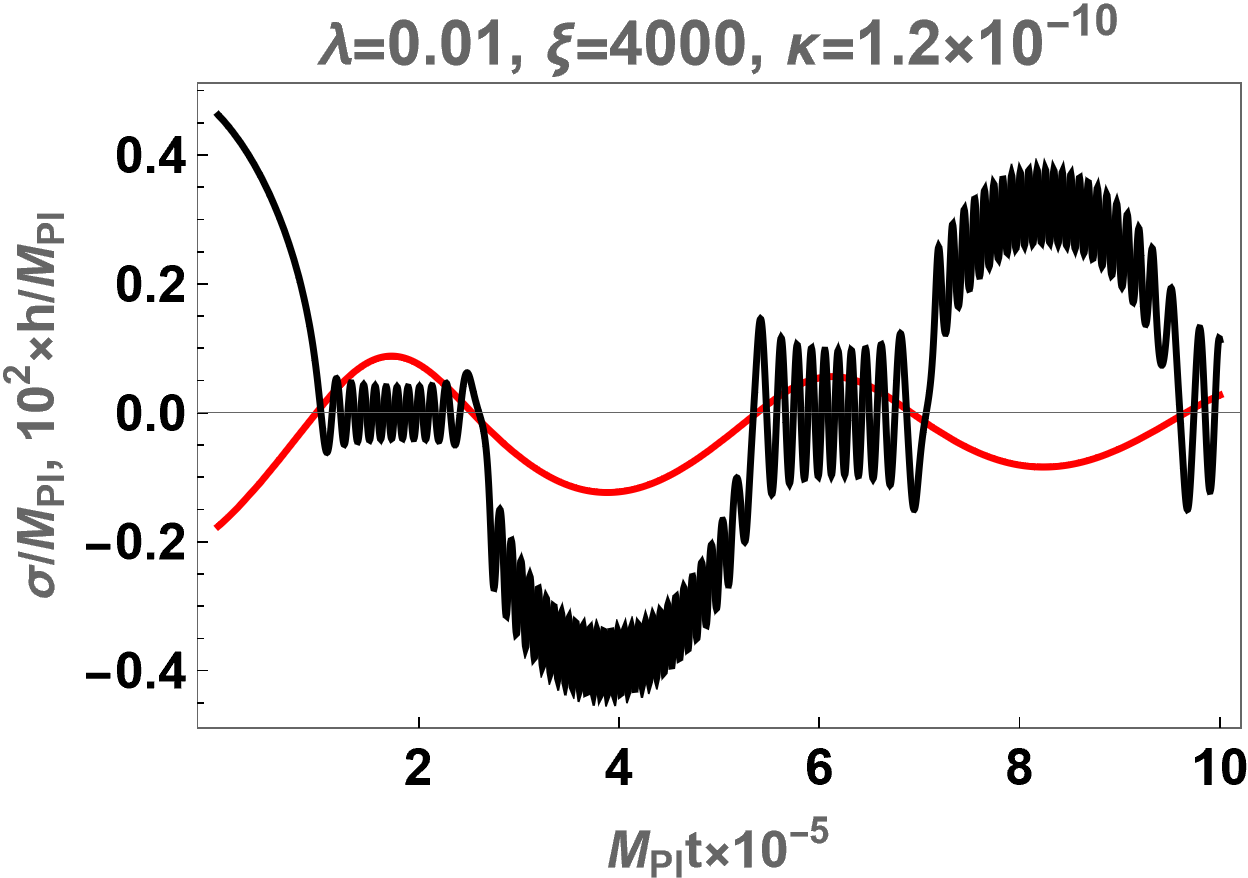}
  \end{center}
 \end{minipage}
 \begin{minipage}{0.5\hsize}
  \begin{center}
   \includegraphics[width=70mm]{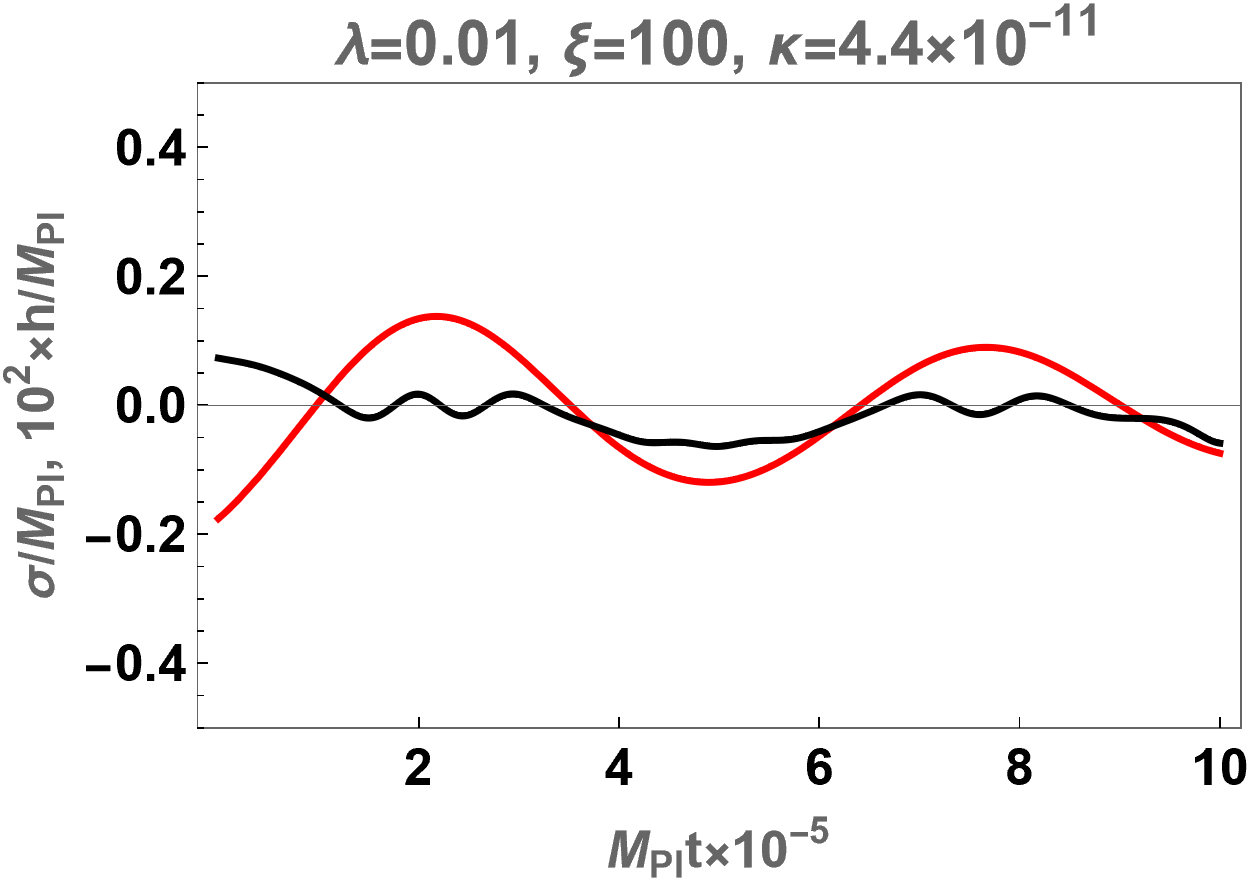}
  \end{center}
 \end{minipage}
\caption{Time evolution of inflaton condensates, $\sigma$ and $h$, during reheating, in red and black lines, respectively. We took $\xi=4000$ ($100$) on left (right) plots.}
  \label{fig:osc}
\end{figure}

Expanding the Lagrangian around $|\sigma|= |\sigma_0|\ll M_{\rm Pl}$ and $h=h_0$, we can read off the masses for $\sigma_0$ and $h_{\rm{osc}}$ as
\begin{align}
m_{\sigma }^{2}= \begin{cases}3\kappa\Mpl^2 \equiv m^2_{\sigma,+} & ,\quad \sigma_0>0, \\ \frac{3\kappa\lambda\Mpl^2}{\lambda+9\kappa\tilde{\xi}^2}\equiv m^2_{\sigma,-}& , \quad \sigma_0<0 \end{cases} \label{m_s}   
\end{align}
and 
\begin{align}
m_{h }^{2}= \begin{cases}3\sqrt{6}\kappa \tilde{\xi} \Mpl\sigma_0\equiv m^2_{h,+} & , \quad\sigma_0>0, \\ 6\sqrt{6}\kappa \tilde{\xi} (-\Mpl\sigma_0)\equiv m^2_{h,-}& , \quad \sigma_0<0,
\end{cases}  \label{m_h}  
\end{align}
respectively.  Thus, the masses of the inflaton condensates are time-dependent, due to the interactions between them. We note that the mass for the sigma condensate for $\sigma_0<0$ in Eq.~(\ref{m_s}) and the masses for the Higgs condensate in Eq.~(\ref{m_h}) are valid for ${\tilde\xi}|\sigma_0|/M_{\rm Pl}\gg 1$. This is true if the inflaton field value is not far from the one at the end of inflation because ${\tilde\xi}|\sigma_e|/M_{\rm Pl}\simeq 20-700\gg 1$ for ${\tilde\xi}=100-4000$. Then, the masses for the inflaton condensates in Eqs.~(\ref{m_s}) and (\ref{m_h}) are good approximations for the perturbative reheating. As will be shown in the next subsection, the Higgs condensate with a relatively large positive $\sigma_0$ dominates the perturbative reheating, so we focus on the regime.

In Fig.~\ref{fig:osc}, we depict the numerical solution to Eqs.~$\eqref{EOM_sigma}$-$\eqref{Friedmann}$ for the background evolution of $\sigma$ and $h$, with the initial condition set by Eq.~$\eqref{chi_e}$. We set $\Gamma_{\sigma_0}=\Gamma_{h_0}=\Gamma_{h_{\rm{osc}}}= 0$ for the early time after inflation.
Thus, we find that the time evolution of the Higgs condensate is well approximated by Eq.~(\ref{hv}) for a large $\tilde{\xi}$, as seen in Fig.~\ref{fig:osc}. But, the deviation of the Higgs condensate from $h_0(\sigma_0)$ has  a large oscillation frequency for $\tilde{\xi}\gtrsim 100$.
The rapidly oscillating part $h_{\rm osc}$ of the Higgs condensate appears prominent for $\sigma_0>0$ when the Higgs background $h_0$ becomes zero. It turns out that $h_{\rm osc}$ is the dominant source for reheating due to the rapid oscillation and the large top Yukawa coupling, as will be discussed shortly.

%%%%%%%%%%%%%%%%%%%%%%%%%%%%%%%%%%%%%%%%%%%%%%%%%%%%%%%%%%

\subsection{Decay rates of inflaton condensates}

We derive the decay rates of the sigma and Higgs condensates. For the analytic approach, we approximate the inflaton condensates to $\sigma_0(t)  \sim \sin (m_{\sigma} t)$ and $ h_{\rm{osc}}(t)\sim \sin(m_h t)$ with constant masses given in Eqs.~$\eqref{m_s}$ and $\eqref{m_h}$, and neglect the expansion of the Universe for the decay rates. However, we need to go beyond such approximations when  the dynamics of $\sigma$ and $h$ becomes nonlinear and far from the harmonic oscillator\footnote{A more rigorous treatment has been made in Ref.~\cite{He:2020qcb}}. 

We first divide $\sigma$ and $h$ into the inflaton condensates and the quantum fluctuation parts, $\delta \sigma $ and $\delta h$, as follows, 
\begin{align}
&\sigma=\sigma_0(t)+\delta \sigma,\label{def_fl_s} \\ &h=h_0(\sigma_0)+h_{\rm{osc}}(t)+\delta h. \label{def_fl_h}    
\end{align}

%%%%%%%%%%%%%%%%%%%%%%%%%%%
\subsubsection*{Decay rates of the sigma condensate}

Expanding the Lagrangian~$\eqref{L_E}$ by Eqs.~$\eqref{def_fl_s}$ and $\eqref{def_fl_h}$, we find the dominant terms for  the $\sigma_0$ decay as
\bea
\mathcal{L}\supset c\sigma_0(\delta h)^2
\eea
with
\bea
c= \begin{cases}-\frac{3}{2}\sqrt{6}\kappa\tilde{\xi}\Mpl  & , \quad\sigma_0>0, \\ 3\sqrt{6}\kappa\tilde{\xi}\Mpl& , \quad\sigma_0<0.\end{cases}    
\eea
$\sigma_0$ couples only to the other particles in the Standard Model through the conformal factors, $\Delta$ and $\Omega^2$, and thus suppressed by the Planck scale. 
Applying the standard formula for the decay rate of the inflaton condensate~\cite{Ichikawa:2008ne,Nurmi:2015ema,Kainulainen:2016vzv}, we obtain  
\bea
\Gamma_{\sigma_{0} \rightarrow \delta h\delta h} 
= \begin{cases}\frac{9\sqrt{3}}{16 \pi}\Mpl\kappa^{3/2}\tilde{\xi}^2\left(1-4\sqrt{6}\tilde{\xi}\frac{\sigma_0}{\Mpl}\right)^{1/2}  & ,\,\,\sigma_0>0, \\ \frac{9\sqrt{3}}{4 \pi}\Mpl\kappa^{3/2}\tilde{\xi}^2\sqrt{\frac{\lambda_{\rm eff}}{ \lambda}}\left(1+8 \sqrt{6}\tilde{\xi}\frac{\lambda_{\rm eff}}{\lambda} \frac{\sigma_{0}}{\Mpl}\right)^{1 / 2}& ,\,\,\sigma_0<0. \end{cases}  \label{Gamma_s}  
\eea
In either cases, $\sigma_0>0$ or $\sigma_0<0$, for a sizable non-minimal coupling with $\tilde{\xi}\gtrsim 1$, the $\sigma_{0} \rightarrow \delta h\delta h$ decay mode is kinematically blocked in the early stage of reheating, but it is limited for $|\sigma_0|/M_P\lesssim 0.1\tilde{\xi}^{-1}$ after some oscillations, being subdominant for reheating.

\subsubsection*{Decay rates of the Higgs condensate}

As discussed in the previous section, the Higgs condensate is composed of the slowly oscillating part $h_0$ related to $\sigma_0$ and the relatively rapidly oscillating part $h_{\rm osc}$. 

First, regarding the decays of $h_{\rm osc}$ that starts appearing for $\sigma_0>0$, we focus on the decay mode into a top quark pair through the Yukawa coupling~$y_t$. From the following interaction for the Higgs condensate,
\begin{align}
\mathcal{L}\supset -\frac{y_t}{\sqrt{2}}h\bar{t}t=-\frac{y_t}{\sqrt{2}}(h_0(\sigma_0)+h_{\rm{osc}})\bar{t}t, \label{int_h_t}
\end{align}
we identify the effective top quark mass as
\begin{align}
m_t=y_t\sqrt{\frac{3\sqrt{6}\kappa}{2\lambda_{\rm eff}}\tilde{\xi} (-\Mpl\sigma_0)},   
\end{align}
for $\sigma_0<0$, but $m_t=0$ for $\sigma_0>0$. From Eq.~$\eqref{int_h_t}$, we also obtain the decay rate for $h_{\rm osc}\to t{\bar t}$ as
\begin{align}
\Gamma_{h_{\rm{osc}} \rightarrow t \bar{t}}= \begin{cases} \frac{3 y_{t}^{2}}{16 \pi}\Mpl\left(3 \sqrt{6}\kappa \tilde{\xi}\frac{\sigma_0}{\Mpl}\right)^{1 / 2}  & , \quad\sigma_0>0, \\ \frac{3 y_{t}^{2}}{16 \pi}\Mpl\left(-6 \sqrt{6}\kappa \tilde{\xi}\frac{\sigma_0}{\Mpl}\right)^{1 / 2} \left(1-\frac{y_t^2}{\lambda_{\rm eff}}\right)^{3 / 2}& ,\quad \sigma_0<0. \end{cases}  \label{Gamma_h}  
\end{align}
Here, for $\sigma_0<0$, the $h_{\rm osc}\to t{\bar t}$ decay mode is kinematically allowed, only if the non-minimal coupling $\tilde{\xi}$ is large enough, $\tilde{\xi}\gtrsim 5000$ for $y_t=0.5$ at inflation scale. However, for $\sigma_0>0$, the $h_{\rm osc}\to t{\bar t}$ decay mode is always open~\cite{He:2020qcb}, thus it becomes a dominant decay mode for the Higgs condensate. Similarly, the other decay modes of  $h_{\rm osc}$ such as $h_{\rm osc}\to WW, ZZ, b{\bar b}$ can be open (for large $\tilde{\xi}$ in the case of gauge bosons) \cite{He:2020qcb}, but they are subdominant as compared to $h_{\rm osc}\to t{\bar t}$.

On the other hand, the slowly oscillating part of the Higgs condensate, $h_0$, has a nonzero amplitude only for $\sigma_0<0$, with a characteristic frequency of order $m_\sigma$. Thus, the decay mode for $h_0\to t{\bar t}$ is open for $|\sigma_0|/M_P\lesssim 0.1( \lambda/y^2_t) \tilde{\xi}^{-1}$, with the corresponding decay rate  given by $\Gamma_{h_0\to t{\bar t}}\sim y^2_t m_\sigma$. In this case, $h_0\to t{\bar t}$ is kinematically blocked in the wider field range of $\sigma_0$ for  $\lambda \ll y^2_t$ than for $\sigma_{0} \rightarrow \delta h\,\delta h$.  

To conclude, for most of the field range of the sigma condensate with $\tilde{\xi} |\sigma_0|/M_{\rm Pl}\gtrsim 1$, the decay modes for the sigma condensate and the $h_0$ part of the Higgs condensate are kinematically blocked, while the $h_{\rm osc}$ part has a larger decay rate from $h_{\rm osc}\to t{\bar t}$ for $\sigma_0>0$ and it becomes a dominant source for reheating.

\subsection{Analytic and numerical solutions for reheating} \label{RT_MT}

In this subsection, we solve  the Boltzmann equations with the decay rates derived in the previous subsection, and study the evolution of inflaton and radiation energy densities, the reheating and maximum temperatures, and the equation of state.

\subsubsection*{Analytic solutions}

We first derive the analytical solutions for energy densities during reheating.   
Our system contains two inflatons, $\sigma$ and $h$, which makes it difficult to follow the dynamics analytically\footnote{The reheating analysis in multifield inflation model has been discussed in Refs.~\cite{Battefeld:2008bu,Choi:2008et,Battefeld:2009xw,Braden:2010wd,Meyers:2013gua,Elliston:2014zea,Hotinli:2017vhx,Leung:2012ve,Huston:2013kgl,Leung:2013rza,Watanabe:2015eia,DeCross:2015uza,DeCross:2016cbs,DeCross:2016fdz,Schimmrigk:2017jwa,Gonzalez:2018jax,Martin:2021frd}.}. As mentioned in Sec.~\ref{BG_after_inflation}, the evolution of both inflaton condensates are intertwined through nonlinear interactions and affected by anharmonic terms.  
But, in order to capture the essence of the reheating dynamics, we split  the total energy density and pressure into $\rho_{\sigma+h}\simeq \rho_{\sigma}+\rho_{h}$ and $p_{\sigma+h}\simeq p_{\sigma}+p_{h}$, respectively, and treat them in the separate Boltzmann equations. Here, we note that
\begin{align}
&\rho_{\sigma}=\frac{1}{2}\dot{\sigma}^2+\frac{1}{2}m^2_{\sigma,+}\sigma^2, \ \  \rho_{h}= \frac{1}{2}\dot{h}^2+\frac{1}{2}m^2_{h,+} h^2+\frac{\lambda_{\rm{eff}}}{4}h^4,\\
&p_{\sigma}=\frac{1}{2}\dot{\sigma}^2-\frac{1}{2}m^2_{\sigma,+}\sigma^2, \ \  p_{h}= \frac{1}{2}\dot{h}^2-\frac{1}{2}m^2_{h,+} h^2-\frac{\lambda_{\rm{eff}}}{4}h^4,
\end{align}
where $m^2_{\sigma,+}$ and $m^2_{h,+}$ are given by
\begin{align}
&m^2_{\sigma,+}=3\kappa \Mpl^2, \label{ms1} \\
&m^2_{h,+}=3\sqrt{6}\kappa \tilde{\xi} \Mpl\sigma_0.  \label{mh1}
\end{align}
Here, the masses for the inflaton condensates in Eqs.~(\ref{ms1}) and (\ref{mh1}) are taken from Eqs.~(\ref{m_s}) and (\ref{m_h}) for $\sigma_0>0$. But, when $\sigma_0$ becomes negative during the oscillation, the Higgs condensates becomes tachyonic, developing a nonzero VEV very quickly and switching to $m^2_{\sigma,-} $and $m^2_{h_-}$ for $\sigma_0<0$ as shown in Eqs.~(\ref{m_s}) and (\ref{m_h}). 

\begin{figure}[t]
 \begin{minipage}{0.5\hsize}
  \begin{center}
   \includegraphics[width=70mm]{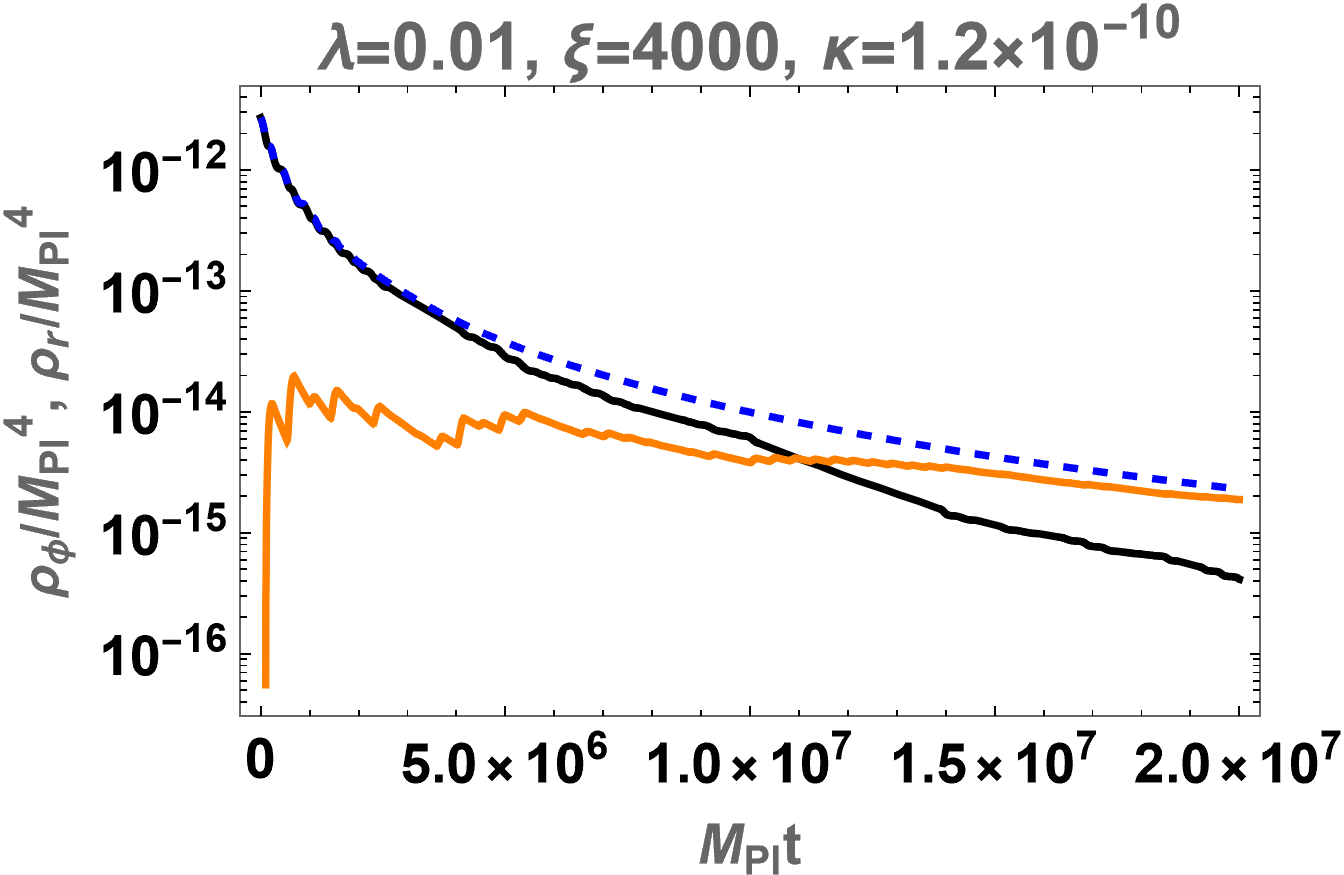}
  \end{center}
 \end{minipage}
 \begin{minipage}{0.5\hsize}
  \begin{center}
   \includegraphics[width=70mm]{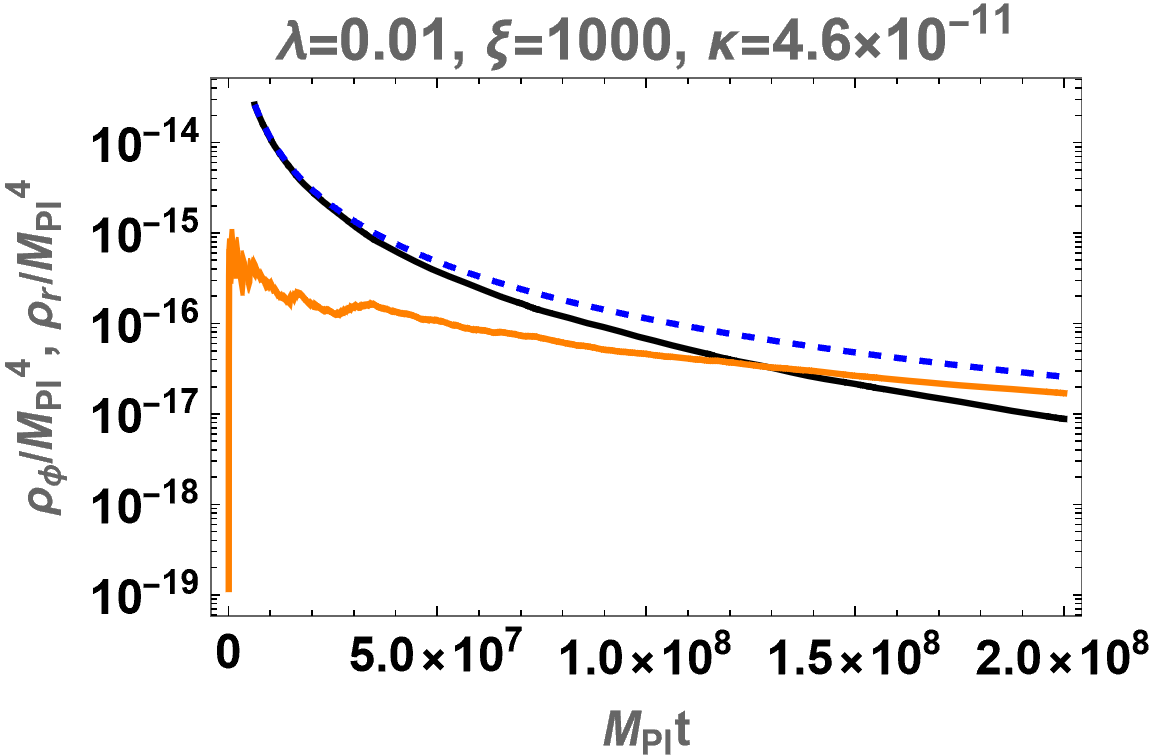}
  \end{center}
 \end{minipage}
 
 \begin{minipage}{0.5\hsize}
  \begin{center}
   \includegraphics[width=70mm]{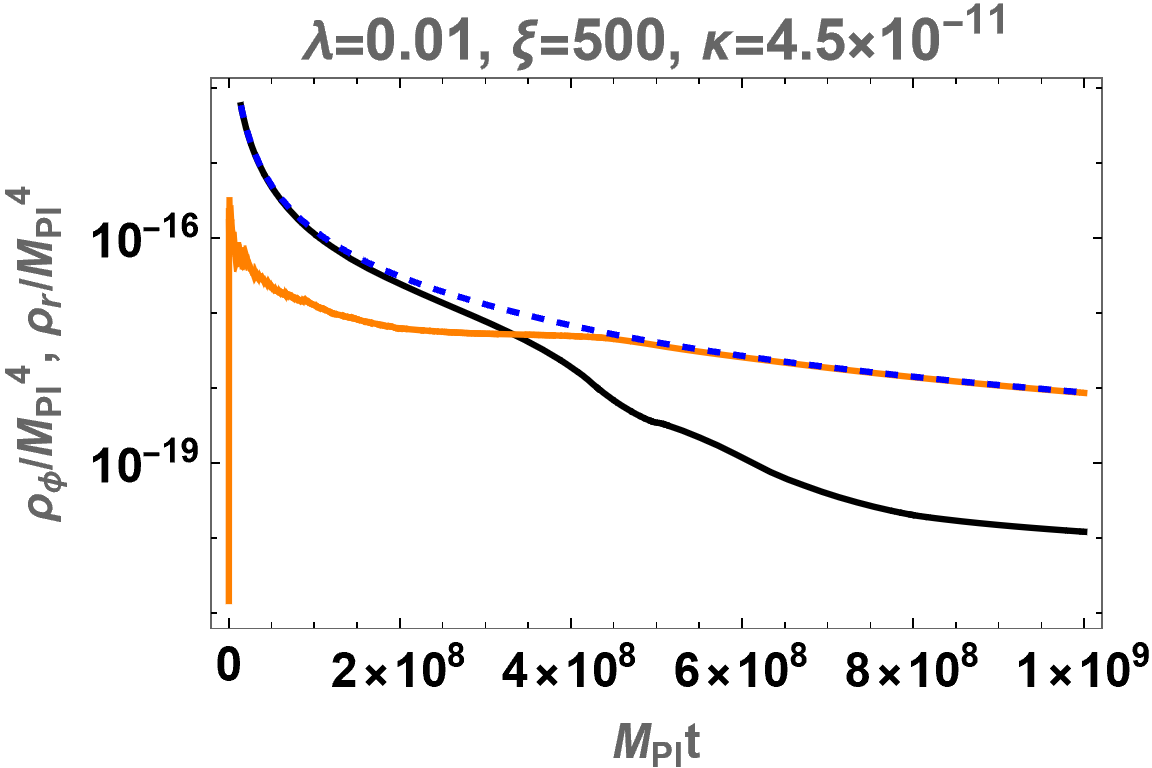}
  \end{center}
 \end{minipage}
 \begin{minipage}{0.5\hsize}
  \begin{center}
   \includegraphics[width=70mm]{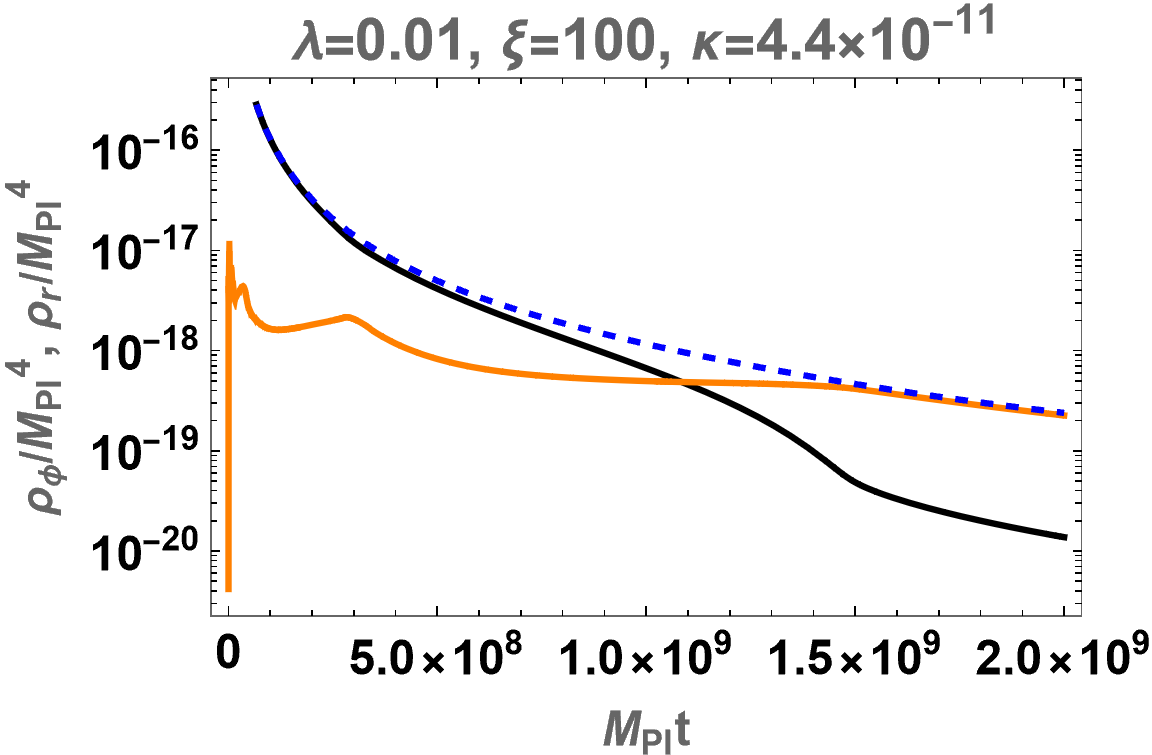}
  \end{center}
 \end{minipage}
 
 \caption{Time evolution of  the energy densities during reheating. Inflaton energy density, $\rho_\phi\equiv \rho_{\sigma+h}$, the radiation energy density, $\rho_r$, and the total energy density, $\rho_{\sigma+h}+\rho_r$, are shown in black, orange and blue dashed lines, respectively. We chose $\xi=4000, 1000, 500, 100$, for top left, top right, down left, and down right plots, respectively.}
  \label{fig:rhophi_vs_rhor}
\end{figure}

Using $\rho_{\sigma,h}$ and $p_{\sigma,h}$, we recast the set of Boltzmann equations, Eqs.~$\eqref{EOM_sigma}$-$\eqref{Friedmann}$, into the following,
\begin{align}
&\dot{\rho}_{\sigma}+3H(\rho_{\sigma}+p_{\sigma})+\Gamma_{\sigma}(\rho_{\sigma}+p_{\sigma})=0,\label{CE_sigma}\\
&\dot{\rho}_{h}+3H(\rho_{h}+p_{h})+\Gamma_{h}(\rho_{h}+p_{h})=0,\label{CE_h}\\
&\dot{\rho}_r+4H\rho_r-\Gamma_{\sigma}(\rho_{\sigma}+p_{\sigma})-\Gamma_{h}(\rho_{h}+p_{h})=0,\label{Rapp}\\
&3\Mpl^2H^2=\rho_{\sigma}+\rho_{h}+\rho_r \label{H2app}
\end{align}
where we omitted the higher order terms suppressed by the Planck scale.

Before going into the details on reheating, we comment on the initial energy densities at the onset of inflaton oscillations.
At the end of inflation with $\epsilon=1$, we recall $\sigma_e\simeq -0.18M_{\rm Pl}$ and  $h_e$ is given by Eq.~(\ref{hv}) with $\sigma=\sigma_e$. Thus, we have the inflaton condensates as ${\dot\sigma}^2_e=\frac{1}{2}m^2_{\sigma,-} \sigma^2_e$  and ${\dot h}^2_e\simeq h^2_e\cdot\frac{{\dot \sigma}^2_e}{4\sigma^2_e}=\frac{1}{8} m^2_{\sigma,-} h^2_e$ at the end of inflation. 
Then, we find the sum of the energy density and pressure at the end of inflation:
\bea
\rho_\sigma + p_\sigma &\simeq&  \frac{1}{2}m^2_{\sigma,-} \sigma^2_e, \\
\rho_h + p_h &\simeq& \frac{1}{8} m^2_{\sigma,-} h^2_e.
\eea
with $m^2_{\sigma,-}=\frac{3\lambda \kappa}{\lambda_{\rm eff}}\, M^2_{\rm Pl}$.
Here, we note that the ratio of the inflatons at the end of inflation is given by
\bea
\frac{h^2_e}{\sigma^2_e}\simeq \frac{3\sqrt{6} \kappa \tilde{\xi}}{0.18 \lambda_{\rm eff}}. 
\eea
In the perturbative regime satisfying the CMB normalization, we have $\kappa\tilde{\xi}\lesssim 0.02\lambda_{\rm eff}$, for which $\sigma_e\gtrsim h_e $, so  $\rho_\sigma + p_\sigma \gtrsim \rho_h + p_h$ at the onset of the inflaton oscillation. 
But, as discussed in the previous subsection, in most of the field values of $\sigma_0$, the decays of the inflaton condensates are kinematically blocked, due to the large effective masses of the decay products, so they are not efficient for reheating.

Nonetheless, for $\sigma_0>0$, the $h_{\rm osc}$ part of the Higgs condensate starts appearing and always decays by $h_{\rm osc}\to t{\bar t}$ with a large decay width as discussed in the previous subsection. Taking $h_{\rm osc}(t)=A\, \cos(m_{h,-} t)$ and an approximate conservation of the Higgs energy density by $\rho_h\sim m^2_{\sigma,-} h^2_e\sim A^2 m^2_{h,-}$, we obtain the initial amplitude for $h_{\rm osc}$ as $A\sim (m_{\sigma,-}/m_{h,-})h_e$. Thus, even if the $h_{\rm osc}$ part has a small amplitude for $m_{h,-}\gg m_{\sigma,-}$, it can reheat the Universe efficiently with the initial Higgs energy at the end of inflation.

We now discuss the approximate solutions for the energy densities during reheating. Assuming that the average of each pressure vanishes during reheating, namely, $p_{\sigma}=p_h=0$, and neglecting $\Gamma_{\sigma} $ and $\Gamma_{h} $ in Eqs.~$\eqref{CE_sigma}$ and $\eqref{CE_h}$, we obtain 
\begin{align}
\rho_{\sigma}=\rho_{\sigma,\rm{end}}\left(\frac{a}{a_{\rm{end}}}\right)^{-3} ,\ \ \rho_{h}=\rho_{h,\rm{end}}\left(\frac{a}{a_{\rm{end}}}\right)^{-3}  \label{inflatonc}
\end{align}
where $a$ is the scale factor, and the subscript ``end" means that the quantities are evaluated at the end of inflation.  Substituting Eq.~(\ref{inflatonc}) into Eq.~$\eqref{Rapp}$ and now including the decay rates for the inflaton condensates, we can solve $\rho_r$ as a function of $a$ as 
\begin{align}
\rho_{r}=\frac{2\sqrt{3}}{5}\Mpl \frac{\Gamma_{\sigma}\rho_{\sigma,\rm{end}}+\Gamma_{h}\rho_{h,\rm{end}}}{\sqrt{\rho_{\rm{end}}}}\left(\left(\frac{a}{a_{\text {end }}}\right)^{-\frac{3}{2}}-\left(\frac{a}{a_{\text {end }}}\right)^{-4}\right),   \label{sol_rhor} 
\end{align}
where $\rho_{\rm{end}}\equiv \rho_{\sigma,\rm{end}}+\rho_{h,\rm{end}}$ is the total energy density of inflatons at the end of inflation and $\rho_r=\frac{\pi^2g_{\rm{reh}}}{30}T^4$.

\begin{figure}[t]

  \begin{center}
   \includegraphics[width=90mm]{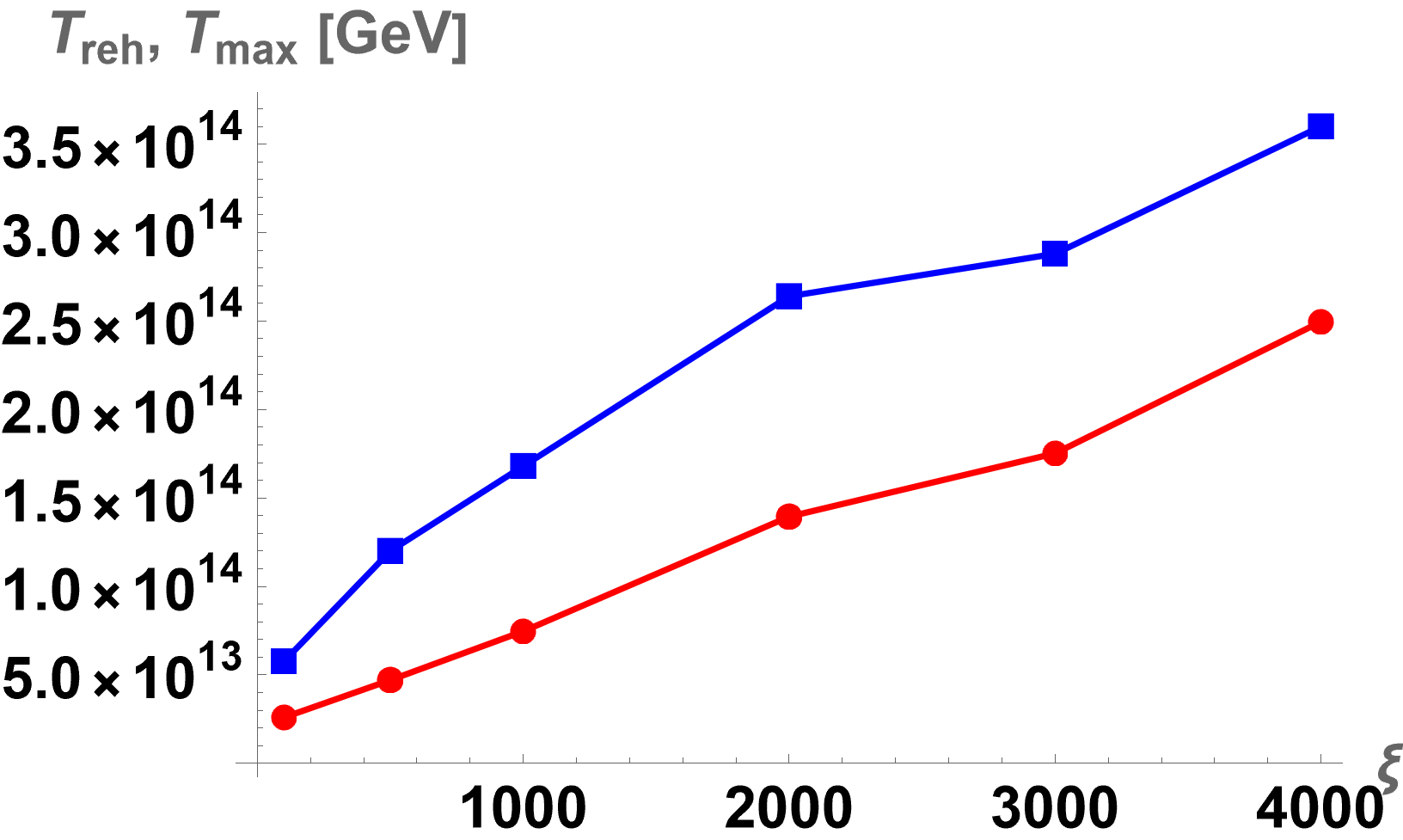}
  \end{center}
 \caption{Reheating temperature $T_{\rm{reh}}$ and maximum temperature $T_{\rm{max}}$ as a function of the Higgs non-minimal coupling $\xi$, in red and blue lines, respectively. }
  \label{fig:sum_T}
\end{figure}

Defining the point of reheating completion at $a_{\rm{reh}}$ by $\rho_{\sigma}(a_{\rm{reh}})+\rho_{h}(a_{\rm{reh}})=\rho_{r}(a_{\rm{reh}})$, we can determine $a_{\rm{reh}}$ by
\begin{align}
\left(\frac{a_{\rm{reh}}}{a_{\rm{end}}}\right)^3= \frac{25}{12}\frac{\rho_{\rm{end}}^3}{\Mpl^2\left(\Gamma_{\sigma}\rho_{\sigma,\rm{end}}+\Gamma_{h}\rho_{h,\rm{end}}\right)^{2}},   
\end{align}
where we used $a_{\rm{reh}}\gg a_{\rm{end}}$. 
Then, the reheating temperature $\rho_{r}(a_{\rm{reh}})=\frac{\pi^2g_{\rm{reh}}}{30}T_{\rm{reh}}^4$ can be expressed as
\begin{align}
T_{\rm{reh}}^4= \frac{72\Mpl^2}{5\pi^2g_{\rm{reh}}}\left(\frac{\Gamma_{\sigma}\rho_{\sigma,\rm{end}}+\Gamma_{h}\rho_{h,\rm{end}}}{\rho_{\sigma,\rm{end}}+\rho_{h,\rm{end}}}\right)^2.    
\end{align}

As the radiation energy from Eq.~$\eqref{sol_rhor}$ is maximized at $a_{\rm{max}}=(8/3)^{2/5}\,a_{\rm{end}}$, we can obtain the analytic expression for the maximum temperature $T_{\rm{max}}$ from
\begin{align}
T_{\rm{max}}^4=   \frac{12\sqrt{3}}{\pi^2g_{\rm{reh}}}\left(\frac{3}{8}\right)^{\frac{3}{5}} \Mpl\frac{\Gamma_{\sigma}\rho_{\sigma,\rm{end}}+\Gamma_{h}\rho_{h,\rm{end}}}{\sqrt{\rho_{\rm{end}}}}.
\end{align}
The above results are a two-field generalization of the previous results~\cite{Chung:1998rq,Giudice:2000ex,Ellis:2015pla,Ellis:2015jpg,Garcia:2017tuj}.

\begin{figure}[t]
 \begin{minipage}{0.5\hsize}
  \begin{center}
   \includegraphics[width=70mm]{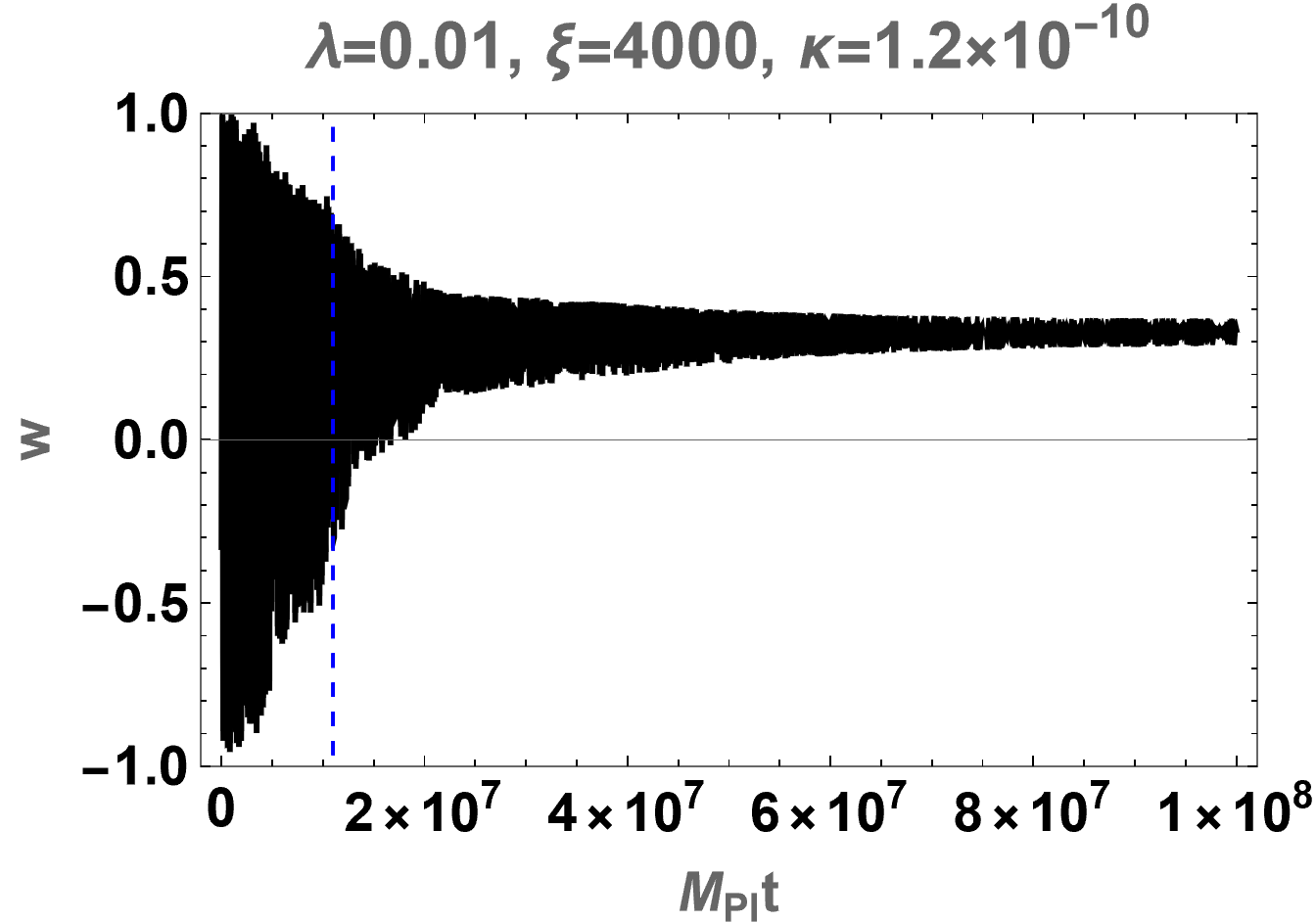}
  \end{center}
 \end{minipage}
 \begin{minipage}{0.5\hsize}
  \begin{center}
   \includegraphics[width=70mm]{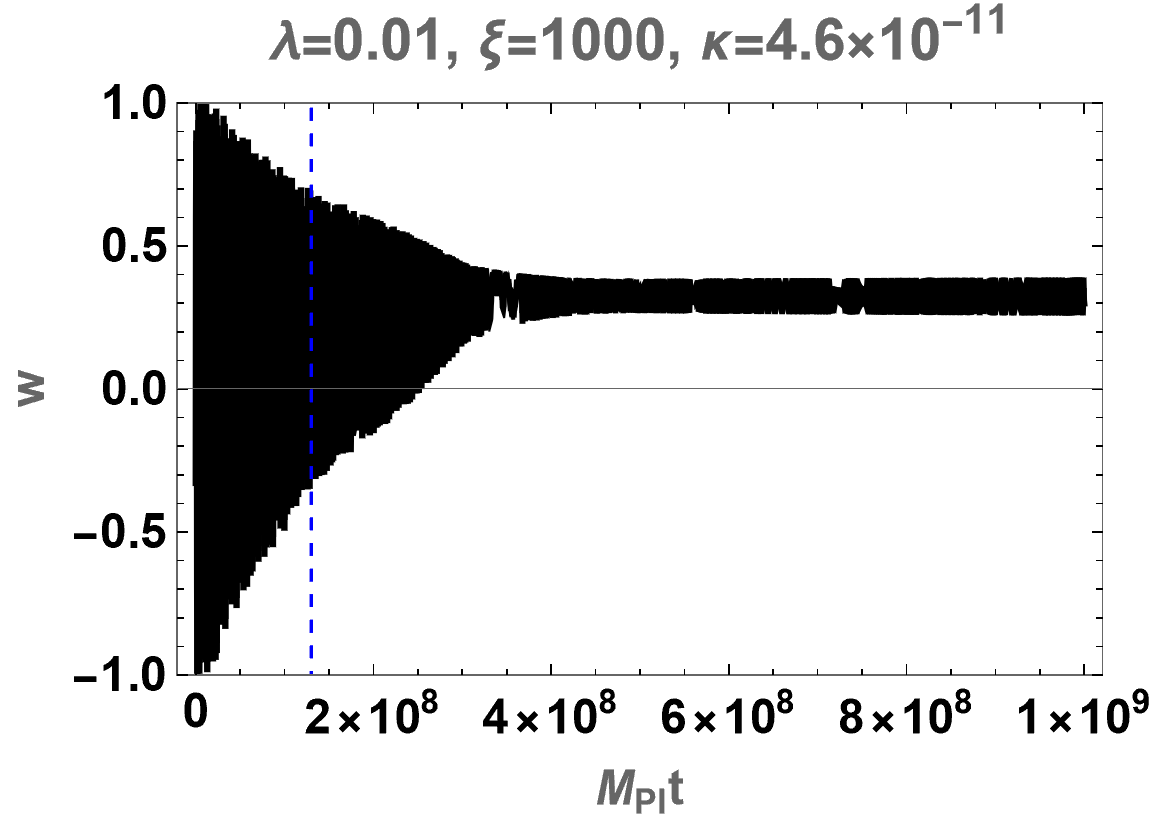}
  \end{center}
 \end{minipage}
 
 \begin{minipage}{0.5\hsize}
  \begin{center}
   \includegraphics[width=70mm]{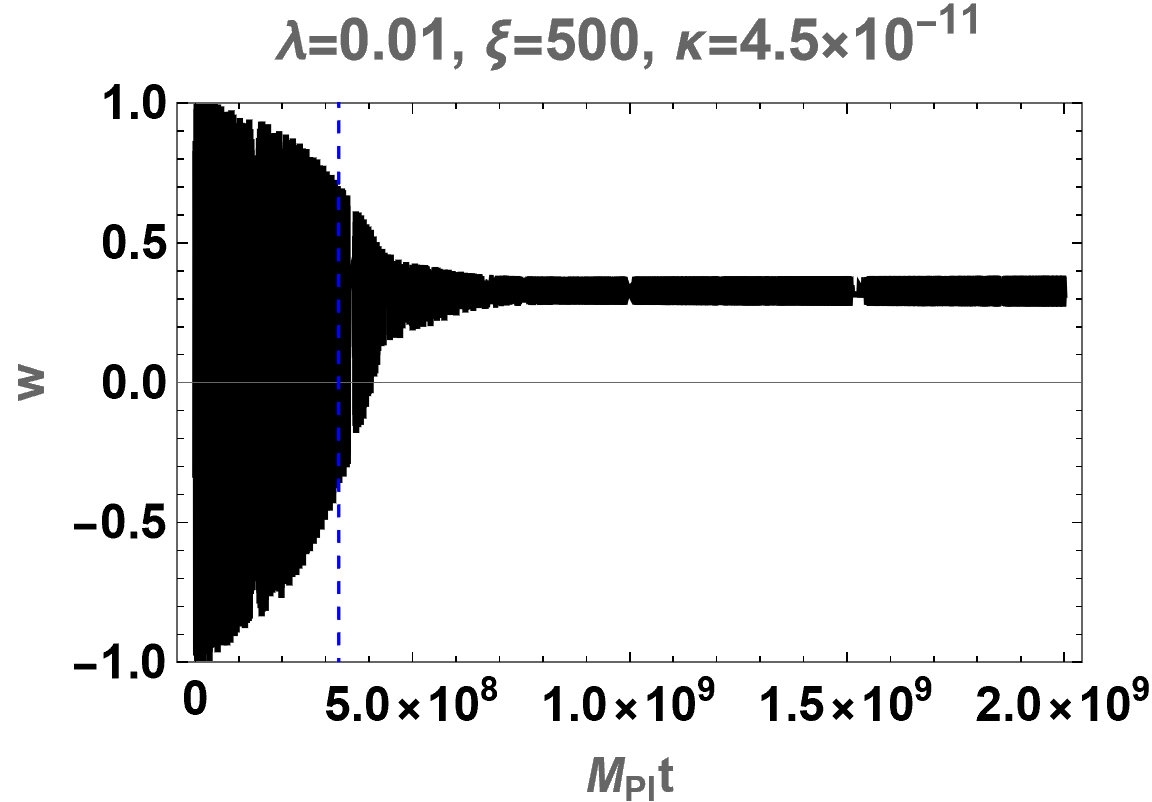}
  \end{center}
 \end{minipage}
 \begin{minipage}{0.5\hsize}
  \begin{center}
   \includegraphics[width=70mm]{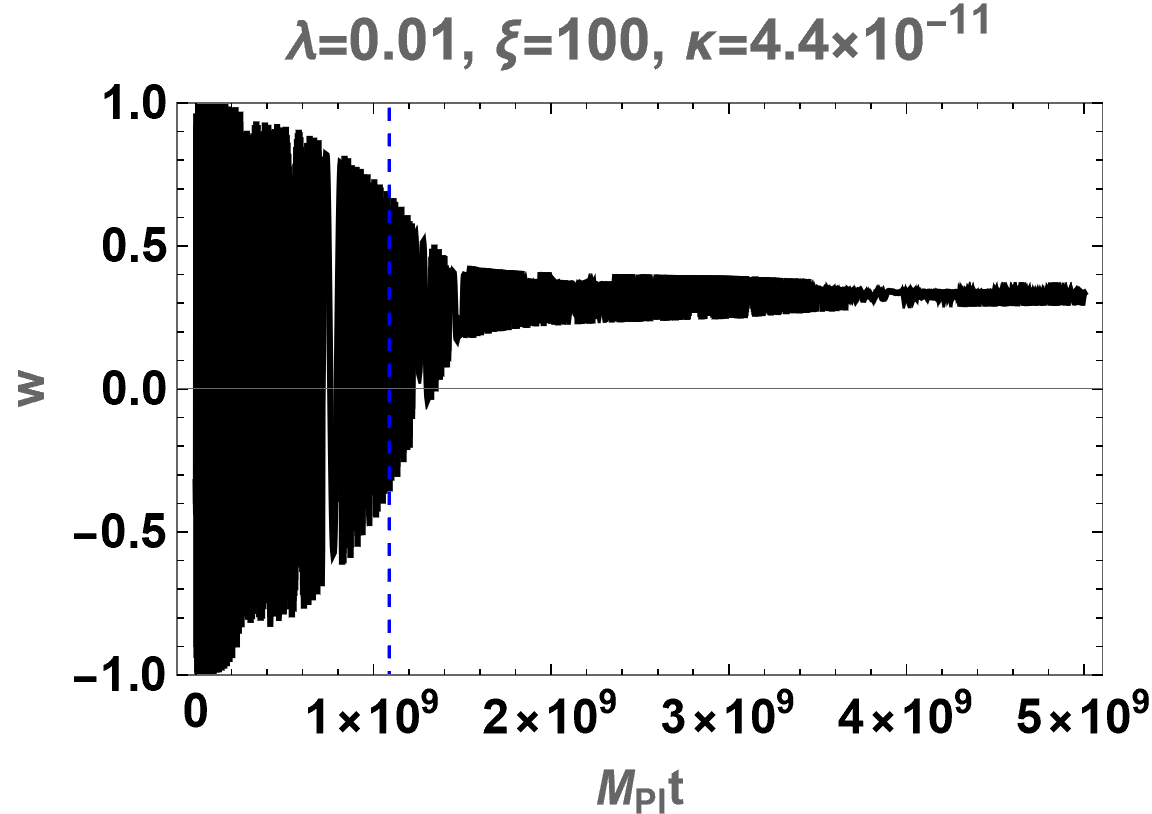}
  \end{center}
 \end{minipage}
 
 \caption{Time evolution of the equation of state $w$ during reheating. The blue dashed line denotes the time at reheating completion. We took the Higgs non-minimal coupling to $\xi=4000, 1000, 500, 100$, for top left, top right, down left, and down right plots, respectively.}
  \label{fig:w}
\end{figure}

\subsubsection*{Numerical solutions}

We are now in a position to present the results for the numerical solutions for the inflaton condensates and the radiation plasma produced during reheating.  We choose different parameter sets with the initial condition given by
Eq.~$\eqref{chi_e}$. The decay rates of the inflaton condensates depend on the sign of $\sigma_0$, which is included by the step function. 
In Fig.~\ref{fig:rhophi_vs_rhor}, we show the numerical results for the time evolution of the inflaton energy density, $\rho_{\sigma+h}$ (black), and the radiation energy density, $\rho_r$ (orange), for different values of the Higgs non-minimal coupling, $\xi=4000, 1000, 500, 100$, from top to bottom panels. Another parameter $\kappa$ (which is the inverse coefficient of $R^2$-term) is chosen appropriately to satisfy Eq.~$\eqref{CMB}$ with $\lambda =0.01$. The blue dashed line denotes the total energy density  $\rho_{\sigma+h}+\rho_r$. The reheating completes when $\rho_{\sigma+h}=\rho_r$. As can be seen from all the plots in Fig.~\ref{fig:rhophi_vs_rhor}, the reheating completion is delayed as $\xi$ becomes smaller.

We assume the instantaneous thermalization for radiation and read off the time evolution of temperature just from
$T=\big(\frac{30}{\pi^2g_{\rm reh}}\rho_r\big)^{1/4}$ with $g_{\rm reh}=106.75$ during reheating.
Then, in Fig.~\ref{fig:sum_T}, we depict the reheating temperature $T_{\rm{reh}}$ and the maximum temperature $T_{\rm{max}}$ for different choices of the Higgs non-minimal coupling, $\xi=4000, 1000, 500, 100$, from top to bottom panels. It turns out that that the difference between $T_{\rm{reh}}$ and $T_{\rm{max}}$ is not so significant, not being so sensitive to the change of $\xi$. As a result, for $100\leq \xi\leq 4000$, we find that the reheating temperature is given by $2.6\times 10^{13}\,{\rm{GeV}}\leq T_{\rm{reh}}\leq 2.5\times 10^{14}\,{\rm{GeV}}$, whereas the maximum temperature varies by $5.8\times 10^{13}\,{\rm{GeV}}\leq T_{\rm{max}}\leq 3.6\times 10^{14}\,{\rm{GeV}}$. 

In Fig.~\ref{fig:w}, we also show the numerical results for the time evolution of $w$ during reheating for different values of the Higgs non-minimal coupling, $\xi=4000, 1000, 500, 100$, from top to bottom panels. Then, we find that the average value of $w$ can be well approximated as $\langle w\rangle=0$ (matter-like) until the reheating completes, as denoted in blue dashed line.

Using our numerical results for the reheating temperature $T_{\rm reh}$ and the averaged equation of state, $\langle w\rangle=0$, and the general formula for the number of efoldings in Eq.~(\ref{Nefolds}), we obtain the number of efoldings for the pivot scale, $k=0.05\,{\rm Mpc}^{-1}$, to be in the following range,
\bea
N=53.2-54.0,
\eea
for $T_{\rm reh}=2.6\times 10^{13}-2.5\times 10^{14}\,{\rm GeV}$. Then, from Eq.~(\ref{spectral}), we can determine the spectral index and the tensor-to-scalar ratio as
\bea
n_s&=&0.9608-0.9614,  \\
r&=&0.0041-0.0042.
\eea
Here, we took $V_{\rm end}\simeq V_I\simeq 3M^2_{\rm Pl} H_k^2$. 
Then, our above results are consistent with the Planck 2018 data for the spectral index, $n_s=0.9670\pm 0.0037$ \cite{Planck:2018jri},  and the Planck/BICEP/Keck limit on the tensor-to-scalar ratio at 95\% CL, $r<0.036$ \cite{BICEP:2021xfz}. We find that the delayed reheating gives rise to a correction to the number of efoldings by $-\Delta N=0.88-1.6$, which amounts to $-\Delta n_s=0.00064-0.0012$.

%%%%%%%%%%%%%%%%%%%%%%%%%%%%%%%%%%%%%%%%%%%%%%%%%%%%%%%%%%%%%%%%%%%%%%%%%%%%%%%%%%%%%%%%%%%%%%%%%%%%%%%%%%
%\subsection{Temperature evolution during reheating}

%%%%%%%%%%%%%%%%%%%%%%%%%%%%%%%%%%%%%%%%%%%%%%%%%%%%%%%%%%
\section{Freeze-in dark matter}

In this section, we introduce a singlet scalar dark matter (DM) in the Higgs-$R^2$ inflation model and study the dark matter production from freeze-in processes during and after reheating. 

%%%%%%%%%%%%%%%%%%%%%%%%%%%%%%%%%

\subsection{A model for scalar dark matter}\label{DM_int}

We extend the Higgs-$R^2$ inflation model by adding a scalar dark matter $\hat{X}$ in the Jordan frame Lagrangian,
\begin{align}
\nonumber \mathcal{L} / \sqrt{-g_J}=\ &\frac{1}{2}(\Mpl^2+\xi \hat{h}^2+\eta \hat{X}^2)R_J -\frac{1}{2}(\partial_{\mu} \hat{h})^{2}-\frac{1}{2}(\partial_{\mu} \hat{X})^{2}-\tilde{V}(\hat{h},\hat{X})+\alpha R_J^2+{\cal L}_{\rm{SM}}\\
=\ &\frac{1}{2}(\Mpl^2+\xi \hat{h}^2+\eta \hat{X}^2+4\alpha\hat{\chi})R_J -\frac{1}{2}(\partial_{\mu} \hat{h})^{2}-\frac{1}{2}(\partial_{\mu} \hat{X})^{2}-\tilde{V}(\hat{h},\hat{X})-\alpha\hat{\chi}^2+{\cal L}_{\rm{SM}},
\end{align}
where we recasted the $R^2$ term in terms of the auxiliary field $\hat{\chi}$ in the second line.
Here, $\eta$ is a non-minimal coupling for $\hat{X}$, and $\tilde{V}$ is the scalar potential for $\hat{h}$ and $\hat{X}$, respecting the $Z_2$-symmetry for dark matter, given by
\begin{align}
\tilde{V}(\hat{h},\hat{X})=\frac{\lambda}{4}\hat{h}^4+\frac{m_X^2}{2}\hat{X}^2+\frac{\lambda_X}{4}\hat{X}^4+\frac{\lambda_{hX}}{4}\hat{h}^2\hat{X}^2, \label{V_til}
\end{align} 
and ${\cal L}_{\rm{SM}}$ contains the remaining part of the Standard Model. 

Making a conformal transformation and field redefinitions as in the previous section,
\begin{align}
g_{J\mu\nu}=(\Delta\Omega)^{ -2}g_{E\mu\nu}, \ \ \hat{h}= \Delta h, \ \ \hat{\chi}= \Delta^2 \chi\ \ \hat{X}= \Delta X,   
\end{align}
with 
\begin{align}
&\Delta^{-2} \equiv  \left(1+\frac{\sigma}{\sqrt{6}\Mpl}\right)^{2},\\
&\left(1+\frac{\sigma}{\sqrt{6}\Mpl}\right)^{2}+\xi \frac{h^2}{\Mpl^2}+\eta \frac{X^2}{\Mpl^2}+4\alpha \frac{\chi}{\Mpl^2}=1-\frac{h^2}{6\Mpl^2}-\frac{X^2}{6\Mpl^2}-\frac{\sigma^2}{6\Mpl^2},\\
&\Omega^{ 2}=1-\frac{h^{2}}{6\Mpl^2}-\frac{X^{2}}{6\Mpl^2}-\frac{\sigma^{2}}{6\Mpl^2},
\end{align}
we obtain the following Lagrangian in Einstein frame,
\begin{align}
\nonumber \mathcal{L} / \sqrt{-g_E}=&\ \frac{\Mpl^2}{2}R_E -\frac{1}{2\Omega^4}\left(1-\frac{h^{2}}{6\Mpl^2} -\frac{X^{2}}{6\Mpl^2} \right)\left(\partial_{\mu} \sigma\right)^{2}-\frac{1}{2\Omega^4}\left(1-\frac{\sigma^{2}}{6\Mpl^2} -\frac{X^{2}}{6\Mpl^2} \right)\left(\partial_{\mu} h\right)^{2}\\
\nonumber &-\frac{1}{2\Omega^4}\left(1-\frac{\sigma^{2}}{6\Mpl^2} -\frac{h^{2}}{6\Mpl^2} \right)\left(\partial_{\mu} X\right)^{2}-\frac{h X}{6\Mpl^2\Omega^4}\partial_{\mu}h\partial^{\mu}X-\frac{h \sigma}{6\Mpl^2\Omega^4}  \partial_{\mu} h \partial^{\mu} \sigma\\
&-\frac{ X \sigma}{6\Mpl^2\Omega^4} \partial_{\mu} X \partial^{\mu} \sigma-V+\frac{1}{(\Omega\Delta)^4}{\cal L}_{\rm{SM}},  \label{L_E_DM} 
\end{align}
where the full scalar potential  $V$ is given by
\begin{align}
\nonumber V= &\ \frac{1}{\Omega^4}\Biggl[\frac{1}{4} \kappa\left(\sigma(\sigma+\sqrt{6}\Mpl)+3\tilde{\xi} h^{2}+3\tilde{\eta}X^2\right)^{2}\\
&+\frac{\lambda}{4}h^4+\frac{m_X^2\Delta^{-2}}{2}X^2+\frac{\lambda_X}{4}X^4+\frac{\lambda_{hX}}{4}h^2X^2\Biggr].  \label{V_E_DM}
\end{align}
Here we introduced the notation,
\begin{align}
\tilde{\eta}\equiv \eta+\frac{1}{6}.\label{def_delta}
\end{align}
Then, $\tilde{\eta}=0$ corresponds to the conformal gravity coupling for dark matter. From the Lagrangian~$\eqref{L_E_DM}$ with Eq.~$\eqref{V_E_DM}$, we find that the scalar dark matter $X$ couples feebly to $\sigma$ and $h$ with gravitational interactions, for conformality, $|\tilde{\eta}|\ll 1$, and a vanishing Higgs-portal coupling, $|\lambda_{hX}|\ll 1$.

%%%%%%%%%%%%%%%%%%%%%%%%%%%%%%%%%%%%%%%%%%%%%%%%%%%%%%%%%%%%%%%%%%%%%%%%%%%%%%%%%%%%%%%%%%%%%%%%%%%%%%%%%%%%%%%%%%%%%%%%%%%%%%%%%%%%%%%
\subsection{Dark matter freeze-in after reheating}

Dark matter can be produced in the periods of reheating and post-reheating. The Universe evolves differently in each period, so we consider the freeze-in production of dark matter in both cases separately in the following. Also, there are two kinds of production mechanisms by the SM radiation (thermal production~\cite{Garcia:2017tuj,Chowdhury:2018tzw,Kaneta:2019zgw,Anastasopoulos:2020gbu,Brax:2020gqg,Kaneta:2021pyx}) and the inflaton condensates (non-thermal production~\cite{Ellis:2015jpg,Garcia:2017tuj,Dudas:2017rpa,Kaneta:2019zgw,Garcia:2020eof,Garcia:2020wiy,Mambrini:2021zpp,Clery:2021bwz}). 

In this subsection, we first compute the DM abundance produced after reheating.
In this case,  reheating is complete and the Universe is dominated by the SM radiation, so only the thermal production for dark matter is important.

When dark matter is decoupled from the SM plasma, the DM number density~$n_X$ is governed by the following Boltzmann equation with the production reaction rate~\cite{Hall:2009bx},
\begin{align}
\dot{n}_X+3Hn_X=R(T),\label{DMeq}
\end{align}
where $R(T)$ is the reaction rate for thermal scattering. For the thermal production $i_1(p_1)+i_2(p_2)\rightarrow X(p_3)+X(p_4)$ with the amplitude~$|\mathcal{M}|^2_{i_1+i_2\rightarrow X+X}$, the reaction rate is given by~\cite{Hall:2009bx,Edsjo:1997bg}
\begin{align}
R=\frac{T}{2^{11}\pi^6}\int_{4m_X^{2}}^{\infty} ds \,d\Omega\,  K_{1}\left(\frac{\sqrt{s}}{T}\right)\sqrt{s-4m_X^{2}}\,\overline{\left|\mathcal{M}_{i_1+i_2\rightarrow X+X}\right|^2},\label{R_T}
\end{align}
where $i_{1,2}$ collectively denote the SM radiation, $d\Omega\equiv 2\pi d\cos \theta_{13}$ is the solid angle of momenta formed by ${\bf{p}}_1$ and ${\bf{p}}_3$, and $K_1(z)$ is the first modified Bessel function of the 2nd kind. The overbar in the amplitude means that the symmetric factor of the initial and final states are included.

Using $T\propto a^{-1}$ (hence $\dot{T}=-HT$) and $H=\sqrt{\frac{g_{\rm{reh}}\pi^2}{90}}\frac{T^2}{\Mpl}$ after reheating,
we can rewrite the Boltzmann equation~$\eqref{DMeq}$ as 
\begin{align}
\frac{d Y}{d T}=-\frac{1}{HT^4}R(T)
=-\sqrt{\frac{90}{\pi^{2} g_{\rm{reh}}}}\frac{\Mpl}{T^{6}}R(T),\label{DMeq2}  
\end{align}
where we defined the DM abundance by $Y\equiv n_XT^{-3}$.

\subsubsection*{Thermal production from the contact terms}

After reheating, the inflatons ($\sigma$ and $h$) have stopped oscillation and settled down to the origin. Thus, neglecting VEVs $\sigma_0$ and $h_0$ and denoting the quantum fluctuations $\delta \sigma, \delta h,$ and $\delta X$, simply by  $\sigma, h,$ and $ X$, respectively, we obtain the following type of interactions between DM  and the other particles :
\bea
\mathcal{L}_X/\sqrt{-g}&=&-\frac{X^2}{12\Mpl^2}(\partial_{\mu}\sigma)^2-\frac{X^2}{12\Mpl^2}(\partial_{\mu}h)^2-\frac{h^2}{12\Mpl^2}(\partial_{\mu}X)^2 \nonumber\\
&&-\frac{\sigma^2}{12\Mpl^2}(\partial_{\mu}X)^2  -\frac{X\sigma}{6\Mpl^2} \partial_{\mu}X\partial^{\mu}\sigma
-\frac{h X}{6\Mpl^2} \partial_{\mu}h\partial^{\mu}X \nonumber \\
&&+c_{\sigma XX}\sigma X^2+c_{\sigma\sigma XX}\sigma^2 X^2+c_{h h XX} h^2 X^2+\frac{X^2}{12\Mpl^2}g^{\mu\nu}T^{\rm{SM}}_{\mu\nu},\label{DM_coupling}
\eea
where
\begin{align}
\nonumber &c_{\sigma XX} =-\frac{m^2_X}{\sqrt{6}\Mpl} -\frac{3}{2}\sqrt{6}\kappa \tilde{\eta}\Mpl, \ \ c_{\sigma\sigma XX} =-\frac{m_X^2}{4\Mpl^2}-\frac{1}{2}\kappa( 3\tilde{\eta} +1),\\
&c_{hh XX} =-\frac{m_X^2}{6\Mpl^2}-\frac{9}{2}\kappa \tilde{\xi}\tilde{\eta}-\frac{\lambda_{hX}}{4},\label{c_3}
\end{align}
and $T^{\rm{SM}}_{\mu\nu}\equiv -\frac{2}{\sqrt{-g}}\frac{\delta\left(\sqrt{-g}{\cal L}_{SM}\right)}{\delta g^{\mu\nu}}$ is an energy-momentum tensor of the SM particles with the Higgs contribution being extracted\footnote{In addition to the above interactions, there exist DM self-interactions such as $X^4$ and $X^2(\partial_{\mu}X)^2$, but they are irrelevant for the following discussion.}.
We note that the dark matter couplings coming from $g^{\mu\nu}T^{\rm{SM}}_{\mu\nu}$ vanish in our case because all the SM fermions and gauge bosons are massless during reheating. Therefore, there is no direct coupling between DM and the SM particles at tree level,\footnote{There are nonzero effective couplings between DM and SM gauge bosons by trace anomaly, although they are suppressed by the loop factor \cite{Watanabe:2010vy,Choi:2019osi}.} except for $h$.

Since the reheating is complete at this stage, we only need to consider the thermal production from the SM plasma including $h$ in radiation components (hence we set $m_h=0$). Only the Higgs field in the SM couples directly to DM via the derivative coupling and the mixing quartic coupling in Eq.~$\eqref{DM_coupling}$, resulting in the scattering amplitude for $h+h\rightarrow X+X$, in the following, 
\begin{align}
\mathcal{M}_{h+h\rightarrow X+X}=-\frac{s+2 m_{X}^{2}}{6 M_{\mathrm{Pl}}^{2}}-18 \kappa \tilde{\eta}\tilde{\xi}-\lambda_{h X},   \label{M_h}
\end{align}
where $s$ is the center of mass energy.

\subsubsection*{Thermal production from the graviton exchanges}

Apart from the contact interactions for DM, all the SM particles can interact with DM through
graviton exchanges~\cite{Garny:2015sjg,Garny:2017kha,Tang:2017hvq,Bernal:2018qlk,Barman:2021ugy,Clery:2021bwz,Mambrini:2021zpp,Haque:2022kez,Haque:2021mab}. 
Expanding the metric around flat space $g_{\mu\nu}\simeq \eta_{\mu\nu}+2h_{\mu\nu}/\Mpl$ and  ignoring mixing  quartic terms and higher order terms, we find that
\begin{align}
\mathcal{L}\supset \ &\frac{1}{\Mpl}h^{\mu\nu}\left(T^{\rm{SM}}_{\mu\nu}+T^{h}_{\mu\nu}+T^{\sigma}_{\mu\nu}+T^{X}_{\mu\nu}\right),\label{int_h}
\end{align}
with
\begin{align}
T^{\phi}_{\mu\nu}=\partial_{\mu}\phi\partial_{\nu}\phi-\frac{1}{2}\eta_{\mu\nu}\eta^{\rho\sigma}\partial_{\rho}\phi\partial_{\sigma} \phi -\frac{1}{2}\eta_{\mu\nu}m_{\phi}^2\phi^2, \ \ (\phi=h, \sigma, X). 
\end{align}
Based on the interactions in Eq.~$\eqref{int_h}$, the scalar dark matter $X$ can be produced from the SM plasma. We obtain the scattering amplitude for  $h+h\rightarrow X+X$ with graviton exchanges by
\begin{align}
\mathcal{M}^G_{h+h\rightarrow X+X}=-\frac{1}{\Mpl^2}\frac{(t-m_X^2)(s+t-m_X^2)}{s},    
\end{align}
where $t=\frac{s}{2}\left(\sqrt{1-\frac{4m_X^2}{s}}\cos \theta_{13}-1\right)+m_X^2$ is another Mandelstam variable.
Then, together with Eq.~$\eqref{M_h}$, the total squared amplitude for   $h+h\rightarrow X+X$ is given by
\begin{align}
|\mathcal{M}^{\rm{total}}_{h+h\rightarrow X+X}|^2=\left(\frac{s+2 m_{X}^{2}}{6 M_{\mathrm{Pl}}^{2}}+18 \kappa \tilde{\eta}\tilde{\xi}+\lambda_{h X}+\frac{(t-m_X^2)(s+t-m_X^2)}{s\Mpl^2}\right)^2.\label{M_hGX}
\end{align}
Similarly, the contributions from the other SM particles with graviton exchanges are the following,
\begin{align}
&|\mathcal{M}^G_{f+f\rightarrow X+X}|^2=\frac{-1}{2 \Mpl^{4} s^{2}}\left(s+2 t-2 m^{2}_X\right)^{2}\left(\left(t-m^{2}_X\right)^{2}+s t\right),\label{M_fGX}\\
& |\mathcal{M}^G_{V+V\rightarrow X+X}|^2=\frac{2}{\Mpl^{4} s^{2}} \left(m_{X}^{4}-2 m^{2}_X t+t(s+t)\right)^{2}.\label{M_VGX}
\end{align}

The next job to do is to calculate the reaction rate $R(T)$ from Eq.~$\eqref{R_T}$ with Eqs.~$\eqref{M_hGX}$, $\eqref{M_fGX}$ and $\eqref{M_VGX}$, and perform the integration of the Boltzmann equation~$\eqref{DMeq2}$ based on the $R(T)$. We leave some calculation details for Appendix~\ref{R_detail}, and show the results only.   
After integrating Eq.~$\eqref{DMeq2}$ from $T_{\rm{reh}}$ to $T_*$ with $T_* \ll m_X\ll T_{\rm{reh}}$, we find that the asymptotic value 
of $Y(T_*)$ is fixed independently of $T_*$,
\begin{align}
Y(T_*) \simeq Y(T_{\rm{reh}}) +\frac{\sqrt{10}}{20480\pi^4g_{\mathrm{reh}}^{1 / 2}}\frac{4m_X^4+45\Mpl^4\left(\lambda_{hX}+18\kappa \tilde{\eta} \tilde{\xi}\right)^2}{m_X\Mpl^3}+\frac{209 \sqrt{10}}{240 \pi^{6} g_{\mathrm{reh}}^{1 / 2}} \frac{T_{\mathrm{reh}}^{3}}{M_{\mathrm{Pl}}^{3}}\label{Y_T*}
\end{align}
where $g_{\rm{reh}}$ is treated as constant during the integration. The second (third) term on the right-hand side shows the IR (UV) freeze-in.  We note that the DM abundance at the reheating temperature, $Y(T_{\rm{reh}})$, is to be determined by the initial condition and the dynamics during reheating, which we will discuss below.

%%%%%%%%%%%%%%%%%%%%%%%%%%%%%%%%%%%%%%%%%%%%%%%%%%%%%%%%%%%%%%%%%%%%%%%%%%%%%%%%%%%%%%%%%%%%%%%%%%%%%%%%%%%%%%%%%%
\subsection{Dark matter freeze-in during reheating}

Next we discuss the DM production during reheating for determining $Y(T_{\rm{reh}})$ in Eq.~$\eqref{Y_T*}$. During reheating, the inflaton energy $\rho_{\sigma+h}\simeq \rho_{\sigma}+\rho_{h}$ dominates the Universe, and therefore, the Universe experiences the matter-like epoch with $w=0$ (see Fig.~\ref{fig:w}). 

The temperature $T$ and the scale factor $a$ follow the non-trivial relation~$\eqref{sol_rhor}$. For the DM production during reheating, in general, we need to take into account the production process not only from the SM radiation (via thermal scattering) but also the inflation condensate (via non-thermal scattering). In our case, both  sigma  and Higgs fields are responsible for the non-thermal production, while
the SM radiation is for the thermal production. Thus, we can divide the total DM abundance into thermal and non-thermal contributions, as follows,
\begin{align}
Y(T_{\rm{reh}})=Y_{\rm{thermal}}(T_{\rm{reh}})+Y_{\rm{non-thermal}}(T_{\rm{reh}}).
\end{align}

\subsubsection*{Thermal production from the SM plasma}

We first estimate the thermal production for $Y_{\rm{thermal}}(T_{\rm{reh}})$. In this case, all the SM particles except for the Higgs contribute to the processes with graviton exchanges and thus only Eqs.~$\eqref{M_fGX}$ and $\eqref{M_VGX}$ are relevant. Using $T\propto a^{-3/8} $ (hence $\dot{T}=-\frac{3}{8}HT$) and $H= \sqrt{\frac{\pi^{2} g_{*}}{90}}\frac{T^{4}}{T_{\text {reh }}^{2}}$, the Boltzmann equation~$\eqref{DMeq}$ can be rewritten as 
\begin{align}
\frac{d}{dT}(n_XT^{-8})=-\frac{8}{3HT^9}R(T)= -\frac{8}{3}\sqrt{\frac{90}{\pi^{2} g_{\rm{reh}}}}\frac{\Mpl T_{\rm{reh}}^2}{T^{13}}R(T).   
\end{align}
Integrating the above equation from $T_{\rm{reh}}$ to $T_{\rm{max}}$ and taking the limit of $m_X\ll T_{\rm{reh}}\ll T_{\rm{max}}$ (see Appendix~\ref{R_detail} for details), we obtain
\begin{align}
Y_{\rm{thermal}}(T_{\rm{reh}})\simeq \frac{69\sqrt{10}}{40\pi^6g_{\rm{reh}}^{1/2}} \frac{T_{\rm{reh}}^3}{\Mpl^3}.\label{Y_Treh_th}   
\end{align}

\subsubsection*{Non-thermal production from inflaton condensates}

Next we move to the non-thermal production for $Y_{\rm{non-thermal}}(T_{\rm{reh}})$.  In this case, it is more convenient to use $a$ as the variable to follow the time evolution, instead of $T$~\cite{Clery:2021bwz}. Then, we can rewrite the Boltzmann equation~$\eqref{DMeq}$ as
\begin{align}
\frac{d }{d a}(n_Xa^3)=\frac{a^2R(a)}{H}
\simeq\sqrt{\frac{3}{\rho_{ \rm{end}}}}\Mpl a^2\left(\frac{a}{a_{\rm{end}}}\right)^{3/2}R(a), \label{DMeq1}  
\end{align}
where we used $3\Mpl^2 H^2\simeq \rho_{\sigma}+\rho_{h}$ for inflaton domination era.

For the non-thermal production with the inflaton condensates, the reaction rate can be written~\cite{Clery:2021bwz,Garcia:2020wiy} as
\begin{align}
R=\frac{1}{8\pi}\sum_{n=1}^{\infty}\left|\mathcal{M}_n\right|^2\sqrt{1-\frac{4m_{X,{\rm{eff}}}^2}{n^2\omega^2}},  \label{R_a} 
\end{align}
where $\mathcal{M}_n$ is the transition amplitude for the inflaton condensate with a Fourier mode $n$ and a frequency $\omega$ to the two-particle final state. We note that only the $n=1$ mode contributes to the reaction rate when the inflaton potential during reheating is quadratic. Here, $m_{X,{\rm{eff}}}^2$ is the effective DM mass during reheating, which can be different from the bare mass $m^2_X$ in Eq.~$\eqref{V_til}$.

In contrast to the case for deriving the DM interactions after reheating in Eq.~$\eqref{DM_coupling}$, we now need to keep the VEVs of the inflatons ($\sigma_0$ and $h_{\rm{osc}}$) during reheating. Then, taking a conformal gravity coupling for dark matter, $\tilde{\eta}=0$, and $\lambda_{hX}=0$, we find that the dominant interactions for dark matter come from both the non-derivative couplings given by
\begin{align}
\mathcal{L}\supset \begin{cases}-\frac{\kappa}{2}\sigma_0^2X^2 & , \quad\sigma_{0}>0, \\ -\frac{\kappa}{2}\frac{\lambda}{\lambda+9\kappa \tilde{\xi}^2}\sigma_0^2X^2 & , \quad \sigma_{0}<0,\end{cases}    \label{s_0^2X^2}
\end{align}
and the derivative couplings given by
\begin{align}
\mathcal{L}\supset -\frac{1}{12\Mpl^2}X^2(\partial_{\mu}\sigma_0)- \frac{1}{6\Mpl^2}X\sigma \partial_{\mu}X\partial^{\mu}\sigma_0-\frac{1}{12\Mpl^2}\sigma_0^2(\partial_{\mu}X)\label{der_s_0^2X^2}.
\end{align}
These contact interaction terms contribute to the scattering process, $\sigma_0+\sigma_0\rightarrow X+X$, with the oscillating background~$\sigma_0$. In Eq.~$\eqref{s_0^2X^2}$, the resulting DM interactions do not differ much for $\sigma_0>0$ and $\sigma_0<0$ if $\tilde{\xi}$ is relatively small. Here, the effective mass of DM can be taken to $m_{X,{\rm{eff}}}^2=\kappa \sigma_0^2$ from Eq.~$\eqref{s_0^2X^2}$. The contributions from Eq.~(\ref{s_0^2X^2}) and (\ref{der_s_0^2X^2}) to the scattering amplitude for $\sigma_0+\sigma_0\rightarrow X+X$ are
\begin{align}
&\mathcal{M}_{1}^{\rm{non-der}}=-\frac{\kappa}{4}\sigma_e^2,\label{nt_nd}\\
&\mathcal{M}_{1}^{\rm{der}}=-\frac{\kappa}{8}\sigma_e^2\left(1-\frac{\sigma_0^2}{3\Mpl^2}\right),\label{nt_d}
\end{align}
where $\sigma_{e}$ is the oscillation amplitude of the sigma-field at the end of inflation, which is related to the sigma energy density by $\rho_{\sigma}=\sigma_{e}^2m^2_{\sigma}/2=3\sigma_{e}^2\kappa\Mpl^2/2$.

In addition to the contact interactions, the graviton exchanges with $\eqref{int_h}$ also give rise to non-negligible contributions to $\sigma_0+\sigma_0\rightarrow X+X$~\cite{Mambrini:2021zpp,Clery:2021bwz}, as follows,
\begin{align}
\mathcal{M}_1^{G}=\frac{3}{8} \kappa \sigma_e^{2}\left(1+\frac{\sigma_{0}^{2}}{6 \Mpl^{2}}\right). \label{nt_G}    
\end{align}
Summing up Eqs.~$\eqref{nt_nd}$, $\eqref{nt_d}$, and $\eqref{nt_G}$, we obtain the total scattering amplitude as 
\begin{align}
\mathcal{M}_1^{\rm{total}}= \frac{5}{48}\kappa\sigma_e^2\frac{\sigma_{0}^{2}}{ \Mpl^{2}}.  \label{M_1_total} 
\end{align}
Remarkably, the leading contributions proportional to $\kappa \sigma_e^2$ cancel out, and the resultant total amplitude is suppressed by $\sigma_0^2/\Mpl^2\sim 10^{-2}$ as compared with the contributions only from the contact interactions. 

Finally, from Eq.~$\eqref{R_a}$ with Eq.~$\eqref{M_1_total}$, we obtain the reaction rate for $\sigma_0+\sigma_0\rightarrow X+X$ as
\begin{align}
R_{\rm{scatter}}(a)\simeq \frac{25}{248832 \pi }\frac{\rho_{\sigma}^4}{\kappa^2\Mpl^{12}}=\frac{25}{248832 \pi }\left(\frac{a}{a_{\rm{end}}}\right)^{-12}\frac{\rho_{\sigma,{\rm{end}}}^4}{\kappa^2\Mpl^{12}},\label{R_condensate}
\end{align}
where we extracted the leading term with respect to $\sigma_0^2/\Mpl^2$, and used the averaged value for the sigma condensate $\left\langle\sigma_{0}^{4}\right\rangle=3\sigma_e^4/8=\rho_{\sigma}^2/6\kappa^2\Mpl^4$.

For a vanishing Higgs-portal coupling, $\tilde{\eta}=\lambda_{hX}=0$, the extra contributions coming from the Higgs condensate are always subdominant as compared to those for the sigma condensate discussed above\footnote{We also remark that even for $\tilde{\eta}=0$, there exists a three-point coupling $\sigma_0 X^2$ in addition to Eq.~$\eqref{s_0^2X^2}$, 
\begin{align}
\mathcal{L}\supset \begin{cases}-\frac{m_X^2}{\sqrt{6}\Mpl}\sigma_0X^2 & , \quad \sigma_{0}>0 \\ -\frac{m_X^2}{\sqrt{6}\Mpl}\frac{\lambda-3\kappa\tilde{\xi}+9\kappa\tilde{\xi}^2}{\lambda+9\kappa \tilde{\xi}^2}\sigma_0X^2 & ,\quad \sigma_{0}<0,\end{cases}    \label{s_0X^2}
\end{align}
which leads to the decay of the sigma condensate. The resulting reaction rate is given by
\begin{align}
R_{\rm{decay}}(a)\simeq \frac{1}{72\pi}\frac{m_X^4\rho_{\sigma}}{\kappa\Mpl^4}.    
\end{align}
However, as compared to Eq.~$\eqref{R_condensate}$, the contribution from the sigma decay is small, because
\begin{align}
R_{\rm{decay}}/ R_{\rm{scatter}}\sim \left(\frac{\Mpl}{\sigma_e}\right)^6\left(\frac{m_X}{m_{\sigma}}\right)^4 \ll 1,
\end{align}
for $m_X\ll m_{\sigma}$.}.

Substituting Eq.~$\eqref{R_condensate}$ into the Boltzmann equation in Eq.~$\eqref{DMeq1}$, and integrating between $a_{\rm{reh}}$ and $a_{\rm{end}}$, we finally obtain the DM abundance during reheating at the time of reheating completion,
\begin{align}
Y_{\rm{non-thermal}}(T_{\rm{reh}})\simeq \frac{\sqrt{3}\pi g_{\rm{reh}}}{2239488}\frac{T_{\rm{reh}}}{\kappa^2\Mpl^{11}} \frac{\rho_{\sigma,{\rm{end}}}^4}{\rho_{\rm{end}}^{3/2}}, \label{Y_Treh}
\end{align}
with $n_X(a_{\rm{end}})=0$. Here, we omitted some small factors by taking into account $a_{\rm{reh}}\gg a_{\rm{end}}$.

%%%%%%%%%%%%%%%%%%%%%%%%%%%%%%%%%%%%%%
\subsection{Dark matter abundance}

We are now in the stage to combine out results for the DM abundance both during and after reheating, obtained in the previous subsections.

\subsubsection*{Conformal couplings for dark matter}

We first discuss the full DM abundance when dark matter have conformal couplings. 

Using the asymptotic value of the yield~$\eqref{Y_T*}$ with Eqs.~$\eqref{Y_Treh_th}$ and~$\eqref{Y_Treh}$, the DM relic abundance at present can be determined to be 
\begin{align}
\nonumber\Omega h^2  
=&\ 1.6\times 10^8 \left(\frac{m_X}{1\rm{GeV}}\right)\left(\frac{g_0}{g_{\rm{reh}}}\right)Y(T_*),\\
\nonumber \simeq&\ 1.6\times 10^8 \left(\frac{m_X}{1\rm{GeV}}\right)\left(\frac{g_0}{g_{\rm{reh}}}\right)\Biggl[\frac{\sqrt{3}\pi g_{\rm{reh}}}{2239488}\frac{T_{\rm{reh}}}{\kappa^2\Mpl^{11}} \frac{\rho_{\sigma,{\rm{end}}}^4}{\rho_{\rm{end}}^{3/2}}+ \frac{623\sqrt{10}}{240\pi^6g_{\rm{reh}}^{1/2}}  \frac{T_{\rm{reh}}^3}{\Mpl^3}\\
&+\frac{\sqrt{10}}{20480\pi^4g_{\mathrm{reh}}^{1 / 2}}\frac{4m_X^4+45\Mpl^4\left(\lambda_{hX}+18\kappa \tilde{\eta} \tilde{\xi}\right)^2}{m_X\Mpl^3}\Biggr],\label{Omegah^2}
\end{align}
where $g_0=3.91$ is the number of the effective
relativistic degrees of freedom at present. 

\begin{figure}[t]

  \begin{center}
   \includegraphics[width=90mm]{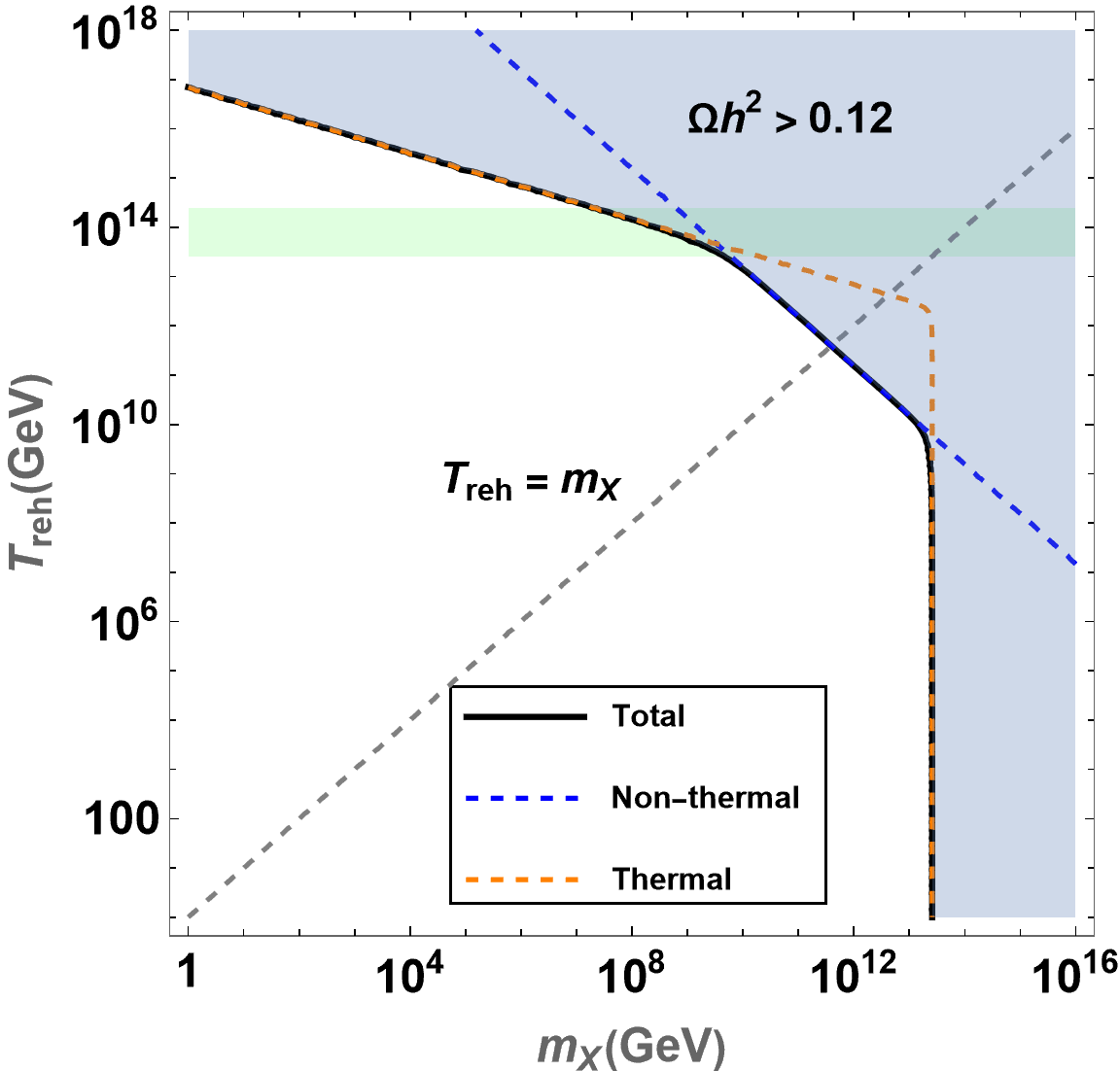}
  \end{center}
 \caption{Parameter space in $(m_X,T_{\rm{reh}})$ for dark matter abundance. The blue shaded region shows $\Omega h^2>0.12$. The green band corresponds to the predicted reheating temperature in the Higgs-$R^2$ model (see Fig.~\ref{fig:sum_T}). The region with $T_{\rm{reh}}>m_X$ in the figure (namely, the region above the gray dashed line) is valid under the current approximation~$\eqref{Y_T*}$ and~$\eqref{Y_Treh_th}$.}
  \label{fig:DM}
\end{figure}

In Fig.~\ref{fig:DM}, we show the parameter space in $(m_X,T_{\rm{reh}})$ for the case with~$\tilde{\eta}=\lambda_{hX}=0$ under the condition that the observed DM abundance is saturated, $\Omega h^2=0.12$. We also set $\rho_{{\rm{end}}}=9\times 10^{61}\,{\rm{GeV}}^4$, which holds almost the same for $100\leq \xi\leq 4000$, and $\rho_{\sigma,{\rm{end}}}=\rho_{{\rm{end}}}$. The orange dashed line in Fig.~\ref{fig:DM} shows the result from the thermal production during and after reheating (from the second and third terms in Eq.~$\eqref{Omegah^2}$), while the blue dashed line shows the one from the non-thermal production during reheating (from the first term in Eq.~$\eqref{Omegah^2}$). The net effect for the DM abundance is shown in black line. The blue shaded region shows the overclosure for dark matter, namely, $\Omega h^2>0.12$. The green band corresponds to the reheating temperature obtained in the Higgs-$R^2$ model as shown in Fig.~\ref{fig:sum_T}, that is, $2.6\times 10^{13}\,{\rm{GeV}}\leq T_{\rm{reh}}\leq 2.5\times 10^{14}\,{\rm{GeV}}$, for $100\leq \xi\leq 4000$. 
We note that only the region with $T_{\rm{reh}}\gg m_X$ in the figure (namely, above the gray dashed line) is consistent with the assumption used in Eqs.~$\eqref{Y_T*}$ and~$\eqref{Y_Treh_th}$.

Therefore, we find that scalar dark matter with mass $2.1\times 10^{7}\,{\rm{GeV}}\leq m_X\leq 4.6\times 10^{9}\,{\rm{GeV}}$ can explain the whole amount of the observed DM abundance in our model. For $m_X< 2.1\times 10^{7}\,{\rm GeV}$, scalar dark matter is less abundant than the observed DM abundance, so we need an extra production mechanism or dark matter.
For the range of DM masses that are consistent with the observed relic density, the velocity of dark matter is sufficiently small at the CMB recombination. For instance, recalling that scalar dark matter is produced dominantly from the inflaton scattering during reheating, we denote the DM velocity by $v_X=p_X/E_X$ with $p_X=m_\sigma (a_{\rm reh}/a_{\rm rec})$ at recombination $a_{\rm rec}$. Then, using $a_{\rm rec}/a_{\rm reh}=T_{\rm reh}/T_{\rm rec}$ and taking $m_\sigma\sim 10^{13}\,{\rm GeV}$ and $T_{\rm reh}\sim 10^{13}\,{\rm GeV}$ in our model, we obtain $v_X\sim {\rm eV}/m_X\sim 10^{-18}$ for $m_X\sim 10^{9}\,{\rm GeV}$, which is small enough to be consistent with the Lyman-$\alpha$ constraint \cite{Bringmann:2016din}.

\subsubsection*{Non-conformal couplings for dark matter}

We comment on the effects of the deviation of the non-minimal coupling from conformality $\tilde{\eta}$, and the tree-level Higgs-portal coupling $\lambda_{hX}$. 

First, for thermal production, $\lambda_{hX}$ and a product of couplings, $ \kappa \tilde{\eta}\tilde{\xi}$, appear additively in Eqs.~$\eqref{Omegah^2}$, so their effects on dark matter production are almost the same. To avoid the overproduction of DM, we need to set the upper limits on them, roughly to $|\lambda_{hX}|\lesssim 10^{-12}$ and $|\tilde{\eta}| \lesssim 10^{-6}$ for $\tilde{\xi} \kappa \sim 10^{-7}$. Around these values, the thermal production with $\lambda_{hX}\neq 0$ or $\tilde{\eta}\neq 0$ stands out to affect the total DM abundance.

Secondly, for non-thermal production, non-zero $\tilde{\eta}$ and $\lambda_{hX}$ lead to the following additional interactions,
\begin{align}
\mathcal{L}\supset \begin{cases}-3 \sqrt{\frac{3}{2}} \Mpl \tilde{\eta} \kappa \sigma_{0} X^{2}   & , \quad\sigma_{0}>0, \\ -3 \sqrt{\frac{3}{2}}\Mpl \kappa \frac{\tilde{\eta} \lambda-\lambda_{hX}\tilde{\xi}/2}{\lambda+9\kappa \tilde{\xi}^2}\sigma_{0}X^2 & , \quad \sigma_{0}<0.\end{cases}    
\end{align}
Taking them for $\sigma_0>0$, we find that the reaction rate from the decay of the inflaton condensate is given by 
\begin{align}
R_{{\rm{decay}}, \tilde{\eta}}(a)\simeq \frac{9}{8\pi}\tilde{\eta}^2 \kappa\rho_{\sigma},  
\end{align}
which is smaller than the one from the inflaton scattering with conformality in Eq.~$\eqref{R_condensate}$. For example, we have $R_{{\rm{decay}}, \tilde{\eta}}/R_{\rm{scatter}}\sim 10^{-2} \ll 1$ for $\tilde{\eta}\sim 10^{-6}$.

\begin{figure}[t]

  \begin{center}
   \includegraphics[width=90mm]{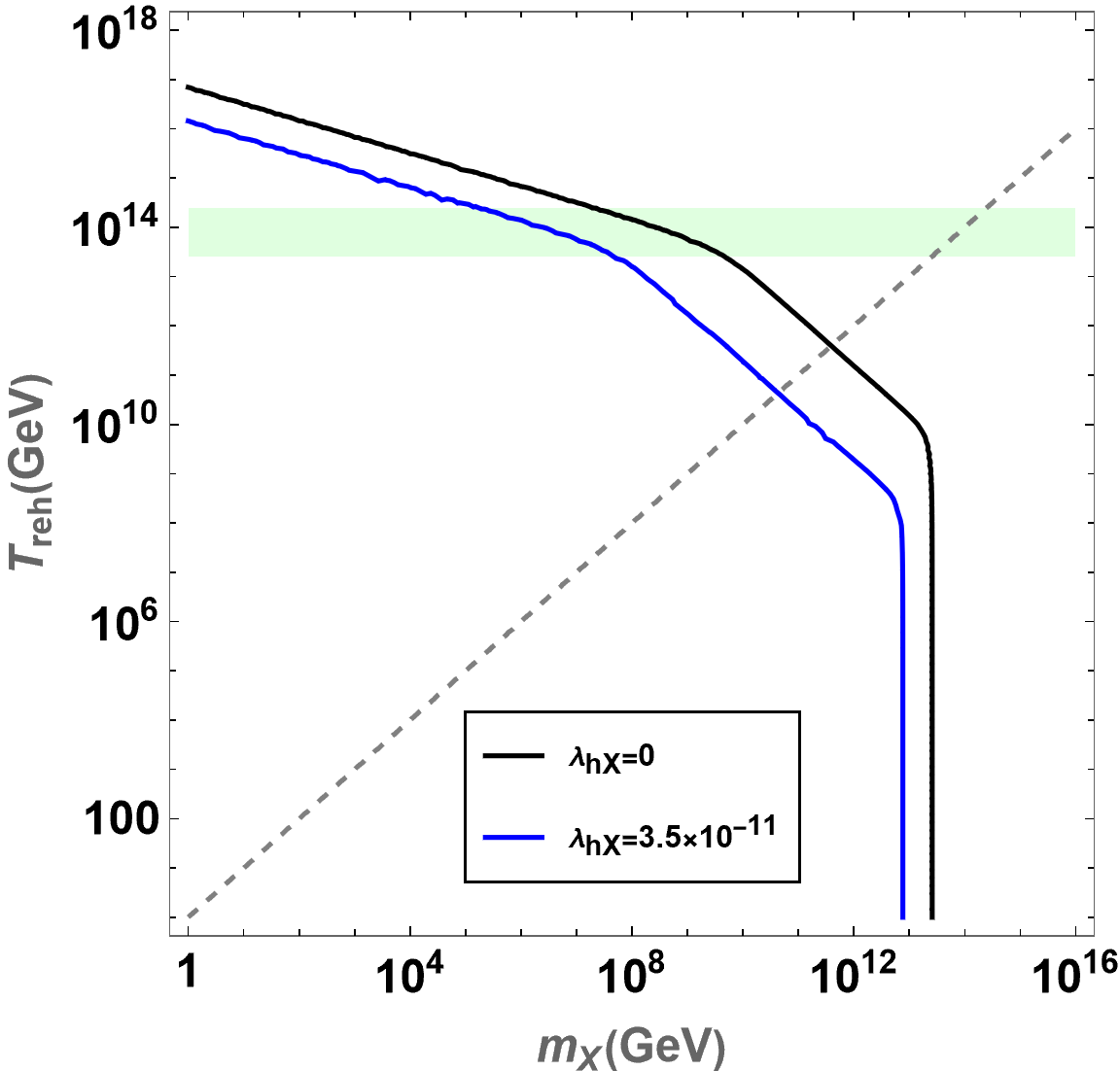}
  \end{center}

 \caption{The same figure as Fig.~\ref{fig:DM}, but the case with a nonzero Higgs-portal coupling, $\lambda_{hX}=3.5\times 10^{-11}$, is included in blue line.}
  \label{fig:DM_lambda}
\end{figure}

In summary, we find that only the thermal production is affected by the non-conformal couplings with $|\lambda_{hX}|\lesssim 10^{-12}$ or/and $|\tilde{\eta}| \lesssim 10^{-6}$. In Fig.~\ref{fig:DM_lambda}, we show the example with $\lambda_{hX}=3.5\times 10^{-11}$. In this case, we can obtain the correct DM relic density for a smaller DM mass than in the case with $\lambda_{hX}=0$.

%%%%%%%%%%%%%%%%%%%%%%%%%%%%%%%%%%%%%%%%%%%%%%%%%%%%%%%%%%%%%%%%%%%%%%%%%%%%%%%%%%%%%%%%%%%%%%%%%%%%%%%%%%%%%%%%%%%%%%%%%%%%%%%%%%%%%%%%%%%%%%%%%%%%%%%%%%

%%%%%%%%%%%%%%%%%%%%%%%%%%%%%%%%%%%%%%%%%%
%%%%%%%%%%%%%%%%%%%%%%%%%%%%%%%%%%%%%%%%%%%%%%%%%%%%%%%%%%
\section{Conclusions}

We have presented the perturbative analysis of reheating dynamics in the Higgs inflation augmented with an $R^2$ term. In this model, there is no unitarity violation from inflation all the way to the end of reheating, as far as perturbativity conditions on the dimensional couplings for the dual sigma and Higgs fields are satisfied. The mixed sigma-Higgs inflation with a large Higgs non-minimal coupling is favored for stability in most of the parameter space, setting the inflaton condensate to be a mixture of sigma and Higgs fields at the onset of oscillations. 

From the perturbative decays of sigma and Higgs condensates, we have identified the evolution of the radiation temperature until the end of reheating in the presence of the perturbative decays of inflaton condensates. Thus, there is no significant delay of reheating completion due to the efficient decays of the Higgs condensate, so the resulting reheating temperature varies between $10^{13}\,{\rm GeV}$ and $10^{14}\,{\rm GeV}$, depending on the non-minimal coupling for the Higgs boson.   

We added a singlet scalar dark matter with a near-conformal gravity coupling and a vanishingly small Higgs-portal coupling in our model and obtained the dark matter relic density from freeze-in processes during and after reheating. 
We found that thermal scattering is most efficient for dark matter production due to  the high reheating temperature of $10^{13}-10^{14}\,{\rm GeV}$ and a correct relic density for dark matter can be obtained for dark matter masses between $10^7\,{\rm GeV}$ and $10^9\,{\rm GeV}$.

%%%%%%%%%%%%%%%%%%%%%%%%%%%%%%%%%%%%%%%%%%%%%%%%%%%%%%%%%%%%%%%%%%%%%%%%%%%%%%%%%%%%%%%%%%%%%%%%%%%%%%%%%%%%%%%%%%%%%%%%%%%%%%%%%%%%%%%%%%%%%%%%%%%%%%%%%%%%%%%%%%%%%%%%%%%%%%%%%%%%%%%%%%

\subsection*{Acknowledgements}
We would like to thank  Dhong-Yeon Cheong, Marcos Garcia, Kunio Kaneta and Yann Mambrini for useful discussion.
HML thanks Yann Mambrini and organizers for invitation to the Paris-Saclay Astroparticle Symposium 2021 during which the current work was still being developed. The work is supported in part by Basic Science Research Program through the National
Research Foundation of Korea (NRF) funded by the Ministry of Education, Science and
Technology (NRF-2022R1A2C2003567 and NRF-2021R1A4A2001897). 
The work of KY is supported by Brain Pool program funded by the Ministry of Science and ICT through the National Research Foundation of Korea(NRF-2021H1D3A2A02038697).

\begin{appendix}

\section{Details on thermal freeze-in}\label{R_detail}
In this Appendix, we show some details on the derivation of  Eqs.~$\eqref{Y_T*}$ and $\eqref{Y_Treh_th} $, which corresponds to thermal production after and during reheating.

Let us start from thermal production after reheating. In this case, we need to evaluate the reaction rate~$\eqref{R_T}$ based on the  scattering amplitudes given in
Eqs.~$\eqref{M_hGX}$, $\eqref{M_fGX}$ and $\eqref{M_VGX}$, which we show here again,
\begin{align}
R(T)=\frac{T}{2^{11}\pi^6}\int_{4m_X^{2}}^{\infty} ds \,d\Omega\,  K_{1}\left(\frac{\sqrt{s}}{T}\right)\sqrt{s-4m_X^{2}}\,\overline{\left|\mathcal{M}_{i_1+i_2\rightarrow X+X}\right|^2},\label{R_T2}
\end{align}
with
\begin{align}
&|\mathcal{M}^{\rm{total}}_{h+h\rightarrow X+X}|^2=\left(\frac{s+2 m_{X}^{2}}{6 M_{\mathrm{Pl}}^{2}}+18 \kappa \tilde{\eta}\tilde{\xi}+\lambda_{h X}+\frac{(t-m_X^2)(s+t-m_X^2)}{s\Mpl^2}\right)^2,\label{M_hGX2}\\
&|\mathcal{M}^G_{f+f\rightarrow X+X}|^2=\frac{-1}{2 \Mpl^{4} s^{2}}\left(s+2 t-2 m^{2}_X\right)^{2}\left(\left(t-m^{2}_X\right)^{2}+s t\right),\label{M_fGX2}\\
& |\mathcal{M}^G_{V+V\rightarrow X+X}|^2=\frac{2}{\Mpl^{4} s^{2}} \left(m_{X}^{4}-2 m^{2}_X t+t(s+t)\right)^{2}.\label{M_VGX2}
\end{align}
Then the total reaction rate can be expressed as
\begin{align}
R_{\rm{total}}(T)=4R_h(T)+45R_f(T)+12R_V(T),  
\end{align}
where $R_h$, $R_f$, and $R_V$ are the reaction rates~$\eqref{R_T2}$ associated with Eqs.~$\eqref{M_hGX2}$, $\eqref{M_fGX2}$, and $\eqref{M_VGX2}$ respectively, with numerical factors corresponding to the SM degrees of freedom. After $\Omega$- and $s$-integrations, we obtain the explicit form 
\begin{align}
\nonumber R_{\rm{total}}(T)=&\ \frac{m_X^2T^2}{5760\pi^5}\frac{4m_X^4+45\Mpl^4\left(\lambda_{hX}+18\kappa \tilde{\eta} \tilde{\xi}\right)^2}{\Mpl^4}  K_1\left(\frac{m_X}{T}\right)^2\\
\nonumber &+\frac{m_X^7 T}{2880 \pi ^{9/2} \Mpl^4}G_{1,3}^{3,0}\left.\left(  \begin{array}{c}
-2 \\
-\frac{7}{2},-\frac{1}{2}, \frac{1}{2}
\end{array}
\right| \frac{m_{X}^{2} }{T^{2}}\right)-\frac{m_X^7 T}{1440 \pi ^{9/2} \Mpl^4}G_{1,3}^{3,0}\left.\left(  \begin{array}{c}
-1 \\
-\frac{5}{2},-\frac{1}{2}, \frac{1}{2}
\end{array}
\right| \frac{m_{X}^{2} }{T^{2}}\right)\\
&+\frac{69 m_X^7 T}{512 \pi ^{9/2} \Mpl^4}G_{1,3}^{3,0}\left.\left(  \begin{array}{c}
0 \\
-\frac{7}{2},-\frac{1}{2}, \frac{1}{2}
\end{array}
\right| \frac{m_{X}^{2} }{T^{2}}\right),
\end{align}
where $K_1(z)$ is the first modified Bessel function of the 2nd kind, and $G_{p, q}^{m, n}\left.\left(\begin{array}{c}
a_{1}, \ldots, a_{p} \\
b_{1}, \ldots, b_{q}
\end{array} \right| z\right)$ is the Meijer G-function.

Next, we integrate the following Boltzmann equation, 
\begin{align}
\frac{d Y}{d T}=-\sqrt{\frac{90}{\pi^{2} g_{\rm{reh}}}}\frac{\Mpl}{T^{6}}R(T),  \label{B_AR}
\end{align}
from $T_{\rm{reh}}$ to a certain late time~$T_*$ ($T_{\rm{reh}}\gg T_*$). Remember that $Y\equiv n_XT^{-3}$. 
The resultant DM abundance at $T=T_*$ is given by
\begin{align}
Y(T_*)=  &\ Y(T_{\rm{reh}}) +\mathcal{F}(T_{\rm{reh}})-\mathcal{F}(T_*)\label{n_exact}
\end{align}
where
\begin{align}
&\mathcal{F}(T)=\mathcal{F}_1(T)+\mathcal{F}_2(T)+\mathcal{F}_3(T)+\mathcal{F}_4(T),\\
&\mathcal{F}_1(T)=-\frac{\sqrt{10}}{7680\pi^{11/2}g_{\rm{reh}}^{1/2}}\frac{4m_X^4+45\Mpl^4\left(\lambda_{hX}+18\kappa \tilde{\eta} \tilde{\xi}\right)^2}{m_X\Mpl^3} G_{2,4}^{3,1}\left.\left(  \begin{array}{c}
1,2 \\
\frac{1}{2},\frac{3}{2}, \frac{5}{2},0
\end{array}
\right| \frac{m_{X}^{2} }{T^{2}}\right),\\
&\mathcal{F}_2(T)=\frac{\sqrt{10}}{1920\pi^{11/2}g_{\rm{reh}}^{1/2}}\left(\frac{m_X}{\Mpl}\right)^3 G_{1,3}^{3,0}\left.\left(  \begin{array}{c}
1\\
-\frac{3}{2},\frac{3}{2}, \frac{5}{2}
\end{array}
\right| \frac{m_{X}^{2} }{T^{2}}\right),\\
&\mathcal{F}_3(T)=\frac{\sqrt{10}}{960\pi^{11/2}g_{\rm{reh}}^{1/2}}\left(\frac{m_X}{\Mpl}\right)^3 G_{2,4}^{3,1}\left.\left(  \begin{array}{c}
1,1\\
-\frac{1}{2},\frac{3}{2}, \frac{5}{2},0
\end{array}
\right| \frac{m_{X}^{2} }{T^{2}}\right),\\&\mathcal{F}_4(T)=-\frac{207\sqrt{10}}{1024\pi^{11/2}g_{\rm{reh}}^{1/2}}\left(\frac{m_X}{\Mpl}\right)^3 G_{2,4}^{3,1}\left.\left(  \begin{array}{c}
1,2\\
-\frac{3}{2},\frac{3}{2}, \frac{5}{2},0
\end{array}
\right| \frac{m_{X}^{2} }{T^{2}}\right).
\end{align}
Here, $g_{\rm{reh}}$ is taken to be constant for $T_*<T<T_{\rm{reh}}$. For $T_{\rm{reh}}\gg T_*$, we find $|\mathcal{F}_1(T_*)|\gg |\mathcal{F}_1(T_{\rm{reh}})|$ but $|\mathcal{F}_{2,3,4}(T_*)|\ll |\mathcal{F}_{2,3,4}(T_{\rm{reh}})|$. This behavior shows that the $\mathcal{F}_1$ contribution corresponds to the IR freeze-in, while those with $\mathcal{F}_{2,3,4}$ are the UV freeze-in. In the limit with~$m_X\gg T_*$, $\mathcal{F}_1(T_*)$ can be approximated as
\begin{align}
\mathcal{F}_1(T_*)\simeq -\frac{\sqrt{10}}{20480\pi^4g_{\rm{reh}}^{1/2}}\frac{4m_X^4+45\Mpl^4\left(\lambda_{hX}+18\kappa \tilde{\eta} \tilde{\xi}\right)^2}{m_X\Mpl^3}.  
\end{align}
As a result,  Eq.~$\eqref{n_exact}$ becomes
\begin{align}
\nonumber Y(T_*)\simeq&\    Y(T_{\rm{reh}})+\frac{\sqrt{10}}{20480\pi^4g_{\rm{reh}}^{1/2}}\frac{4m_X^4+45\Mpl^4\left(\lambda_{hX}+18\kappa \tilde{\eta} \tilde{\xi}\right)^2}{m_X\Mpl^3}\\
&+\mathcal{F}_2(T_{\rm{reh}})+\mathcal{F}_3(T_{\rm{reh}})+\mathcal{F}_4(T_{\rm{reh}}).
\end{align}
We note that the DM abundance at $T=T_*$ is fixed independently of $T_*$.
We can make a further simplification for $T_{\rm{reh}}\gg m_X$, where $\mathcal{F}_{2,3,4}$ can be expanded at the leading order in powers of $m_X/T_{\rm{reh}}$. In this case, we get
\begin{align}
Y(T_*) \simeq Y(T_{\rm{reh}}) +\frac{\sqrt{10}}{20480\pi^4g_{\mathrm{reh}}^{1 / 2}}\frac{4m_X^4+45\Mpl^4\left(\lambda_{hX}+18\kappa \tilde{\eta} \tilde{\xi}\right)^2}{m_X\Mpl^3}+\frac{209 \sqrt{10}}{240 \pi^{6} g_{\mathrm{reh}}^{1 / 2}} \frac{T_{\mathrm{reh}}^{3}}{M_{\mathrm{Pl}}^{3}},
\end{align}
which is same as Eq.~$\eqref{Y_T*}$.

For the estimation during reheating, almost same procedure can apply but this time we need to keep only fermion and gauge boson contributions, Eqs.~$\eqref{M_fGX2}$ and $\eqref{M_VGX2}$ for radiation, and  use the Boltzmann equation during reheating epoch, \begin{align}
\frac{d\tilde{Y}}{dT}= -\frac{8}{3}\sqrt{\frac{90}{\pi^{2} g_{\rm{reh}}}}\frac{\Mpl T_{\rm{reh}}^2}{T^{13}}R(T), \label{B_DR}  
\end{align}  
instead of Eq.~$\eqref{B_AR}$. Here $\tilde{Y}\equiv n_XT^{-8}=YT^{-5}$. The total reaction rate is evaluated as 
\begin{align}
\nonumber R_{\rm{total}}(T)=&\ 45R_f(T)+12R_V(T)\\
=&\ \frac{69m_X^7T}{512\pi^{9/2}\Mpl^4}G_{1,3}^{3,0}\left.\left(  \begin{array}{c}
0\\
-\frac{7}{2},-\frac{1}{2}, \frac{1}{2}
\end{array}
\right| \frac{m_{X}^{2} }{T^{2}}\right).
\end{align}
Then, by integrating Eq.~$\eqref{B_DR}$ from $T_{\rm{max}}$ to $T_{\rm{reh}}$, we obtain
\begin{align}
\tilde{Y}(T_{\rm{reh}}) = \mathcal{G}(T_{\rm{reh}})-\mathcal{G}(T_{\rm{max}})
\end{align}
with 
\begin{align}
\mathcal{G}(T)\equiv \frac{69\sqrt{10}}{128\pi^{11/2}g_{\rm{reh}}^{1/2}}\frac{T_{\rm{reh}}^2}{m_X^4\Mpl^3} G_{2,4}^{3,1}\left.\left(  \begin{array}{c}
1,\frac{11}{2}\\
2,5,6,0
\end{array}
\right| \frac{m_{X}^{2} }{T^{2}}\right).    
\end{align}
We set $\tilde{Y}(T_{\rm{max}})=0$. For $T_{\rm{reh}}\ll T_{\rm{max}}$, we have $|\mathcal{G}(T_{\rm{reh}})|\gg|\mathcal{G}(T_{\rm{max}})|$, and therefore, the asymptotic expression of $Y(T_{\rm{reh}})=T_{\rm{reh}}^5\tilde{Y}(T_{\rm{reh}})$ is fixed by 
\begin{align}
Y(T_{\rm{reh}})\simeq  \frac{69\sqrt{10}}{128\pi^{11/2}g_{\rm{reh}}^{1/2}}\frac{T_{\rm{reh}}^7}{m_X^4\Mpl^3} G_{2,4}^{3,1}\left.\left(  \begin{array}{c}
1,\frac{11}{2}\\
2,5,6,0
\end{array}
\right| \frac{m_{X}^{2} }{T_{\rm{reh}}^{2}}\right).   
\end{align}
In the limit $m_X\ll T_{\rm{reh}}$, it becomes
\begin{align}
Y(T_{\rm{reh}})\simeq \frac{69\sqrt{10}}{40\pi^6g_{\rm{reh}}^{1/2}} \frac{T_{\rm{reh}}^3}{\Mpl^3},
\end{align}
which produces Eq.~$\eqref{Y_Treh_th} $.
%%%%%%%%%%%%%%%%%%%%%%%%%%%%%%%%%%%%%%%%%%%%%%%

\end{appendix}

\end{document}